\documentclass{article}

\usepackage{arxiv}

\usepackage[utf8]{inputenc} 
\usepackage[T1]{fontenc}    
\usepackage{hyperref}       
\usepackage{url}            
\usepackage{booktabs}       
\usepackage{amsfonts}       
\usepackage{nicefrac}       
\usepackage{microtype}      
\usepackage{graphicx}
\usepackage{natbib}
\usepackage{doi}
\usepackage{amsmath}
\usepackage{algorithm}
\usepackage{algpseudocode}
\usepackage{todonotes}
\usepackage{comment}
\usepackage{cancel}
\usepackage{multirow}
\usepackage{xcolor}

\usepackage{listings}
\lstdefinestyle{pythonstyle}{
    language=Python,
    basicstyle=\ttfamily\small,
    keywordstyle=\color{blue}\bfseries,
    stringstyle=\color{red},
    commentstyle=\color{green}\itshape,
    numbers=left,
    numberstyle=\tiny\color{gray}
}

\usepackage{caption}
\usepackage{subcaption}
\usepackage{alphalph}
\usepackage{makecell}



\title{Bi-Level optimization for interpolation-based parameter estimation of differential equations}

\author{ 
    {\hspace{1mm}Siddharth Prabhu}\\
	Department of Chemical \\ and Biomolecular Engineering\\
	Lehigh University\\
	Bethlehem, PA 18015 \\
	\texttt{scp220@lehigh.edu} \\
	\And
    {\hspace{1mm}Srinivas Rangarajan} \\
	Department of Chemical \\ and Biomolecular Engineering\\
	Lehigh University\\
	Bethlehem, PA 18015 \\
	\texttt{srr516@lehigh.edu} \\
        \And
    {\hspace{1mm}Mayuresh Kothare} \\
	Department of Chemical \\ and Biomolecular Engineering\\
	Lehigh University\\
	Bethlehem, PA 18015 \\
	\texttt{mvk2@lehigh.edu} \\
}

\renewcommand{\shorttitle}{}

\begin{document}
\maketitle

\begin{abstract}
    Inverse problem or parameter estimation of ordinary differential equations (ODEs), the iterative process of minimizing the mismatch between model-predicted and experimental states by tuning the parameter values within an optimization formulation, is commonplace in chemical engineering applications. A popular method for parameter estimation is sequential optimization (single-shooting), which numerically integrates the ODE in each iteration. However, computing the gradients for the optimization steps requires calculating sensitivities, i.e., the derivatives of states with respect to the parameters, through the numerical integrator, which can be computationally expensive. In this work, we use interpolation to reduce the cost of these sensitivity calculations. Leveraging this interpolation, we also propose a bi-level optimization framework that exploits the structure of the differential equations and solves a convex inner problem. We apply this framework to examples spanning conventional parameter estimation and the emerging concept of data-driven dynamic model discovery. We show that our approach not only estimates the correct parameters for benchmark problems, but can also be readily extended to delay, stiff, and partially observed differential equations without major modifications.
\end{abstract}


\section{Introduction}

Inverse problem or parameter estimation is a process of obtaining the best parameters of, in this case, ordinary differential equations, given measurements of states. Sequential optimization or single-shooting method formulates this problem as a nonlinear optimization problem \citep{vassiliadis1994solution1, vassiliadis1994solution2}. This optimization problem, with the parameters as the decision variables, minimizes the mean squared error between predicted and measured states. Single-shooting method is difficult to converge for long trajectories, poor initial estimates of the parameters, and highly nonlinear or complex dynamics. Multiple-shooting addresses some of these issues by dividing the trajectory into small intervals, which can be handled in parallel, while adding continuity equations as equality constraints to the optimization problem \citep{bock1984multiple}. This comes at the cost of incorporating states as decision variables, thereby increasing the dimensions of the optimization problem. However, because of the dynamic nature of constraints, the Jacobians and the Hessians of the optimization problem can be computed efficiently. Additionally, the sparse structure of the Hessian can be leveraged by sparse linear solvers, to solve the KKT system more efficiently \citep{Bock1983, doi:10.1137/S1052623401399216, doi:10.1137/S0036144504444711, Andersson2019}. On the other hand, condensing \citep{bock1984multiple, Bock1983} replaces the updates in the states with updates only in the parameters, consequently reducing the dimensionality of Newton-based optimization problems.

These methods, however, require calculating sensitivities \citep{kidger2021on}, which can be computationally expensive. To address this, we use interpolation for parameter estimation as in \citep{bellman1971use, doi:10.1137/0903003}, which is demonstrated only for linearly separable parameters. Interpolation techniques have been used to obtain initial estimates of linearly separable parameters that are sufficiently close to their original value \cite{calver2021using}, which are then used in shooting-based optimization problems. Interpolation has also been used as a computationally cheaper and more stable alternative to computing sensitivities of ordinary differential equations \citep{vajda1986direct}. It is also applied in cascading or bi-level optimization frameworks \citep{ramsay2007parameter}, where, at each iteration, the inner problem fits a linear combination of basis functions or an interpolation function based on the parameters at that iteration, while the outer problem minimizes the squared error between experimental and predicted states using the interpolation function. Alternatively, interpolation can be used to approximate the states in classical collocation methods \citep{biegler2010nonlinear, chen2008efficient}, thereby eliminating the need for numerical integration and recasting the overall optimization problem as one with algebraic equations. However, these methods require that initial state estimates provided by the user be sufficiently close to the true states. A stringent requirement that typically necessitates ad hoc methods to find suitable initial estimates, and therefore requires intervention from domain experts.

\begin{figure}[!ht]
    \centering
    \includegraphics[width = 0.95\linewidth, height = 0.45\textheight, trim = 50 30 50 30, clip]{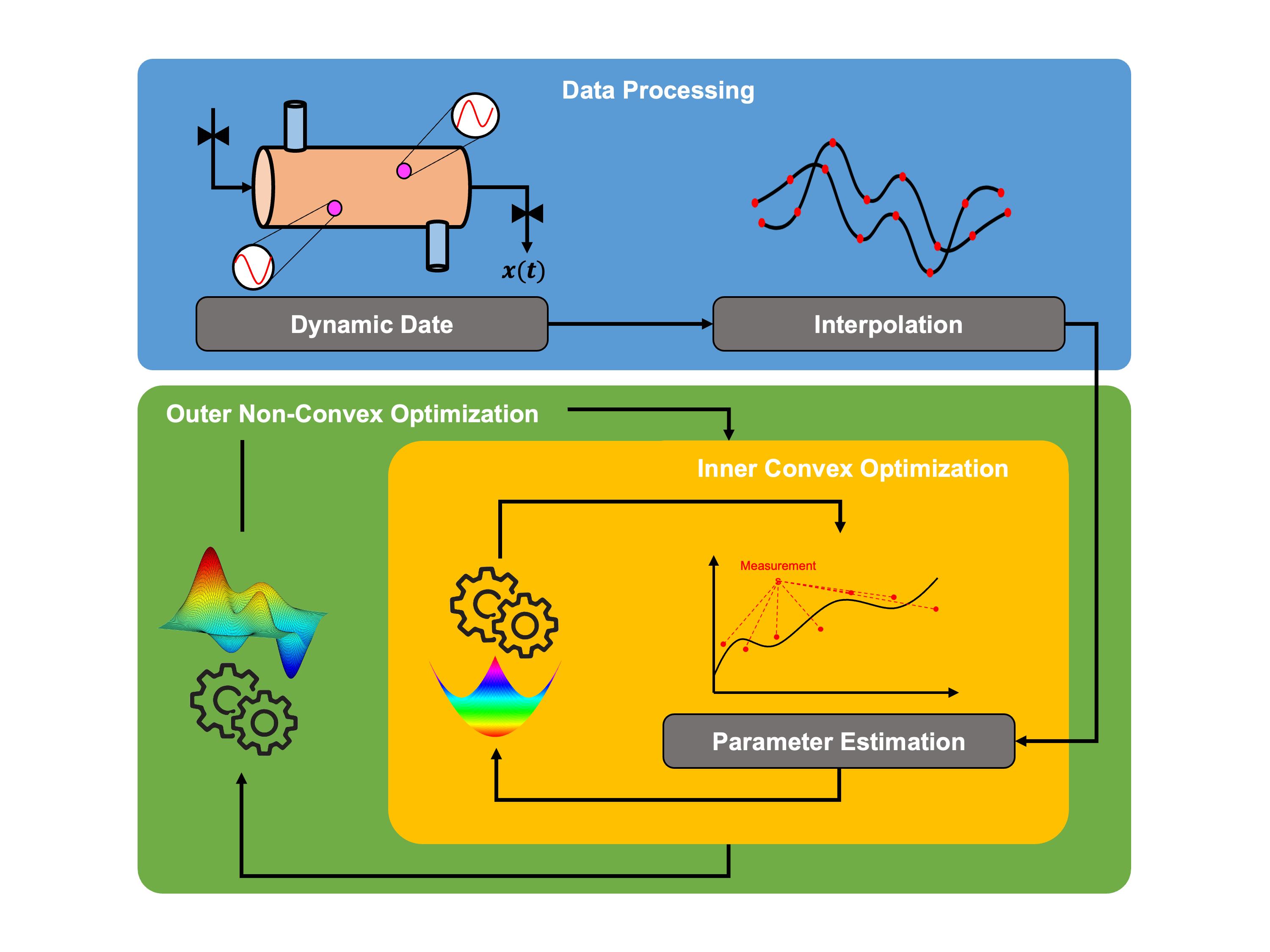}
    \caption{A workflow of the proposed bi-level optimization procedure for parameter estimation of ordinary differential equations}
    \label{fig:overview}
\end{figure}

In our prior work \cite{prabhu2025derivative}, we developed an interpolation-based parameter estimation method that solves a convex optimization problem. While this approach estimates parameters accurately and is data-efficient, it is restricted to differential equations in which the unknown parameters appear linearly. In this work, we extend this approach to a broader class of ordinary differential equations that places no restrictions on how the parameters appear in the system. We formulate a bi-level optimization problem, where the inner problem solves a convex optimization problem in the linear parameters \citep{boyd2004convex}, while the outer problem optimizes over the nonlinear parameters. We use the implicit function theorem \citep{dontchev2009implicit, griewank2008evaluating, 10.1162/neco_a_01547} over the KKT conditions of the inner optimization problem to derive the forward-mode gradient of the optimal solution with respect to nonlinear parameters. An overview of the overall procedure is shown in Figure \ref{fig:overview}. We demonstrate the efficacy of this method on two different yet related classes of benchmark problems. First, we consider parameter estimation for ordinary differential equations (ODE) where the model structure is known but the parameters are unknown. Second, we consider model discovery problems, where both the model structure and parameters of ODE are unknown and are to be learned from the data. Finally, we also demonstrate how our method can be extended to delayed differential equations, stiff differential equations, and partially observed differential equations.

\section{Method}
We extend the problem formulation developed in \cite{prabhu2025derivative} to a broader class of parameter estimation problems of ordinary differential equations. As a result, we consider the following optimization problem  

\begin{align}\label{eqn:sindy}
\begin{split}
    \min _{p, \phi} f(p, \phi) = & \ \frac{1}{2} \sum_{t = t_i}^{t_f} || \hat{X}(t) - \hat{X}(0) - \int_0^t \Theta(X, p, \phi) dt ||^2  \\
    \text{subject to} & \\
    & g(p, \phi) = 0 \\
    & h(p, \phi) \leq 0 \\
    & g'(\phi) = 0 \\
    & h'(\phi) \leq 0
\end{split}
\end{align}

where $\hat{X}(t) \in \mathbb{R}^n$ are the measurements of the states at time $t$, $\Theta(X, p, \phi)$ is the ODE function capturing the temporal dynamics, whose parameters $p$ and $\phi$ need to be estimated. We distinguish between these parameters such that $p$ appears linearly, while $\phi$ appears nonlinearly in the function $\Theta$. As discussed in \cite{prabhu2025derivative}, we approximate the states $X$ as $\Psi(X)(t) : \mathbb{R} \mapsto \mathbb{R}^{n} $ using interpolation, thereby rendering $\Theta$ as only a function of $p, \Phi, t$. We can further separate the linear parameters and integrate as follows 

\begin{equation}
    \int _0^t \Theta(X, p, \phi) dt \approx p \int _0^t \Theta(\Psi(X)(t), \phi) dt
\end{equation}

\begin{figure}[!ht]
    \centering
    \includegraphics[width = \linewidth, height = 0.55\textheight, trim = 25 0 25 0, clip]{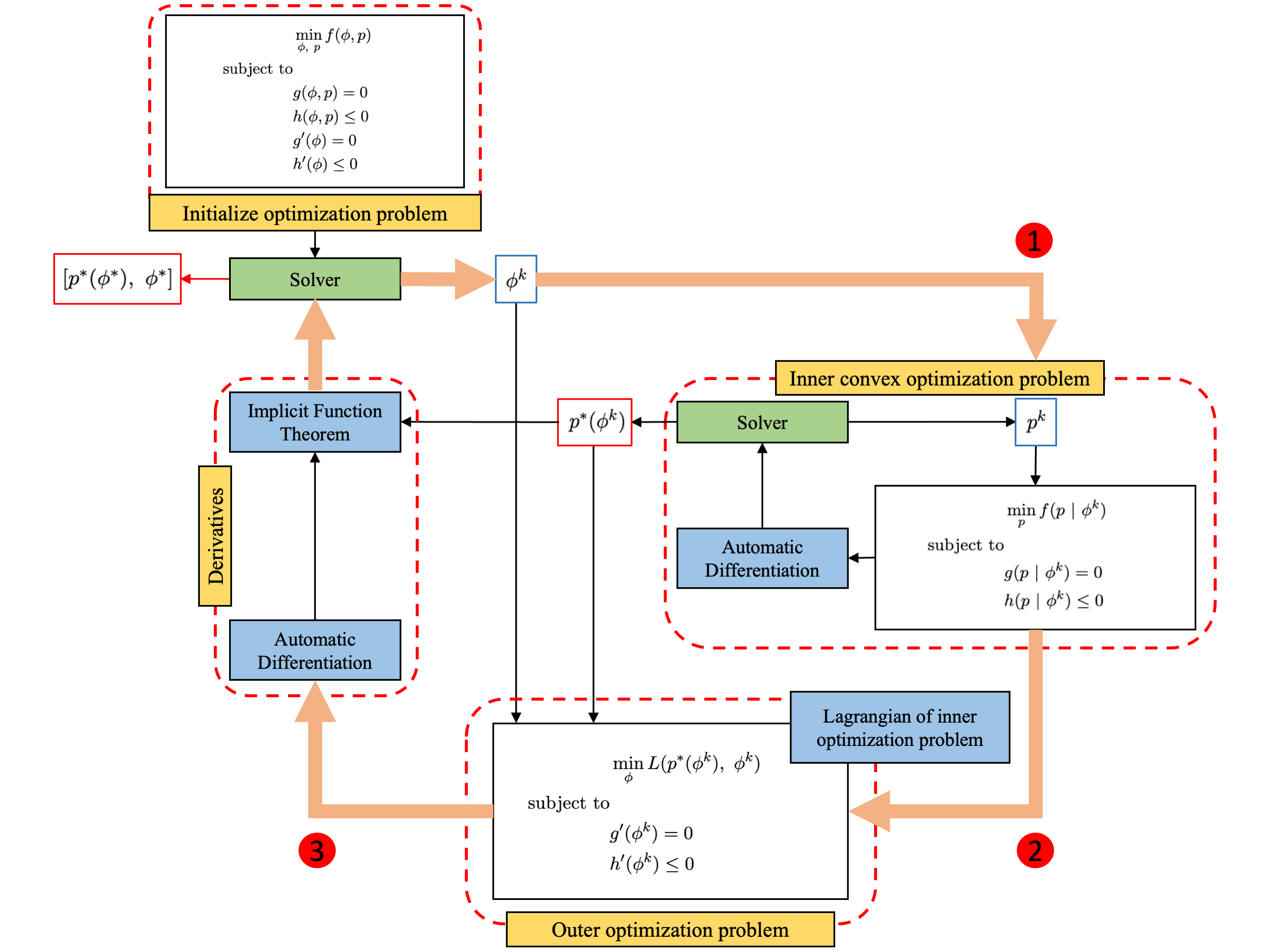}
    \caption{A schematic overview of the optimization problems solved at different levels in the proposed bi-level optimization framework. Gradients computed using automatic differentiation link the inner problem with the outer optimization problem.}
    \label{fig:bilevel}
\end{figure}

Furthermore, we assume that, given $\phi$, the equality constraints, $g \in \mathbb{R}^{m_e}$ are affine in $p$, and the inequality constraints, $h \in \mathbb{R}^{m_i}$, are convex in $p$ \citep{boyd2004convex}. On the other hand, $g'$ and $h'$ can be any nonlinear equality and inequality constraints. Under these assumptions, for a fixed value of $\phi$, the optimization problem becomes convex. We leverage these assumptions and solve a bi-level optimization problem in which the inner problem is convex in $p$ given $\phi$, while the outer problem is non-convex in $\phi$. A schematic overview of the workflow is provided in Figure \ref{fig:bilevel}. At each iteration of the outer optimization, the inner convex problem is first solved using the current value of $\phi$ provided by the solver (step 1 in Figure \ref{fig:bilevel}). Given the resulting optimal $p$ and the current iterate of $\phi$, a nonlinear optimization problem is then solved to get the next update of $\phi$ (step 2 in Figure \ref{fig:bilevel}). The two optimization problems are coupled via gradients computed through automatic differentiation (step 3 in Figure \ref{fig:bilevel}), and this process is repeated until convergence. 

\subsection{Derivative of inner optimization problem}

Since we solve a bi-level optimization problem, we need to find the derivatives across the inner optimization problem. We form the Lagrangian of the inner convex optimization problem as follows

\begin{equation}\label{eqn:lagrangian}
\begin{aligned}
    L(p, \lambda, \mu \ | \ \phi) & = f(p | \phi) + \lambda ^T g(p | \phi) + \mu ^T h(p | \phi)\\
    C(p^*, \lambda ^*, \mu ^*) & = \ \begin{cases}
        f_p(p^* | \phi) + \lambda ^*{}^T g_{p}(p^* | \phi) + \mu ^*{}^T h_p(p^* | \phi) = 0 \quad \text{Stationarity} \\[4pt]
        g(p^* | \phi) = 0 \quad \text{Primal feasibility} \\[4pt]
        \text{diag}(\mu ^*) h(p^* | \phi) = 0 \quad \text{complementary Slackness} 
    \end{cases}
\end{aligned}
\end{equation}

where $\lambda \in \mathbb{R}^{m_e}$ and $ \mu \in \mathbb{R}^{m_i}$ are the Lagrange variables of equality and inequality constraints, respectively. The optimal solution that satisfies Equation \ref{eqn:lagrangian} is given as $(p^*, \lambda ^*, \mu ^*)$. The derivative of the optimal solution with respect to $\phi$ can be efficiently calculated using the implicit function theorem \citep{krantz2002implicit, griewank2008evaluating}. We derive equations used in forward-mode derivative calculations as follows

\begin{equation}
\begin{aligned}
    \frac{d}{d \phi} C(p^* (\phi), \lambda ^* (\phi), \mu ^* (\phi)) & = 0 \\
    \begin{bmatrix}
        \frac{dp ^*}{d \phi} \\[6pt]
        \frac{d\lambda ^*}{d\phi} \\[6pt]
        \frac{d \mu^*}{d \phi}\\[6pt]
    \end{bmatrix} v & = - 
    \begin{bmatrix}
    L_{pp} & g_p^T & h_p^T\\[3pt]
    g_{p} & 0 & 0\\[3pt]
    \text{diag}(\mu) h_p & 0 & \text{diag}(h)
    \end{bmatrix} ^{-1}
    \begin{bmatrix}
        L_{p\phi} \\[3pt]
        g_{\phi} \\[3pt]
        \text{diag}(\mu ^*)h_{\phi}
    \end{bmatrix} v
\end{aligned}
\end{equation}

where $v$ is the tangent vector used in forward-mode automatic differentiation \citep{griewank2008evaluating}. $(g_p, h_p)$ and $(g_{\phi}, h_{\phi})$ are the Jacobian of equality and inequality constraints with $p$ and $\phi$, respectively. $L_{pp}$ is the Hessian of the Lagrangian with respect to $p$. The factorization of the KKT matrix, or the Hessian of the Lagrangian at the optimal point, can be obtained from the optimization solver and reused in derivative calculations. We provide the necessary steps in case the factorization of the KKT matrix is unavailable.

\begin{equation}
\begin{aligned}
    \begin{bmatrix}
        L_{pp} & g_p^T & h_p^T\\[3pt]
        g_{p} & 0 & 0\\[3pt]
        \text{diag}(\mu) h_p & 0 & \text{diag}(h)
    \end{bmatrix} \begin{bmatrix}
        w_1 \\[3pt]
        w_2 \\[3pt]
        w_3
    \end{bmatrix} & = 
    \begin{bmatrix}
        v_1 \\[3pt]
        v_2 \\[3pt]
        v_3
    \end{bmatrix} = - \begin{bmatrix}
        L_{p\phi} v \\[3pt]
        g_{\phi} v \\[3pt]
        \text{diag}(\mu ^*)h_{\phi} v
    \end{bmatrix}
\end{aligned}
\end{equation}

where $v = [v_1, v_2, v_3]^T$ is the Jacobian-vector product of the optimality conditions with the tangent vector $v$ and $w = [w_1, w_2, w_3]^T$ is a vector of appropriate dimensions.

\begin{equation}
\begin{aligned}
    L_{pp} w_1  + g_p^T w_2 + h_p^T w_3 & = v_1 \\
    g_p w_1 & = v_2 \\
    w_3 & = [\text{diag}(h)]^{-1} \left[ v_3 - \text{diag}(\mu) h_p w_1 \right]  
\end{aligned}
\end{equation}

Let $H = \text{diag}(h)$ and $M = \text{diag}(\mu)$ then, substituting for $w_3^T$ gives

\begin{equation}
\begin{aligned}
    \begin{bmatrix}
        L_{pp} - h_p^T M H^{-1} h_p & g_p^T \\[3pt]
        g_{p} & 0 \\
    \end{bmatrix} \begin{bmatrix}
        w_1 \\[3pt]
        w_2 \\
    \end{bmatrix} & =  - 
    \begin{bmatrix}
        v_1 - h_p^T H^{-1} v_3 \\[3pt] 
        v_2
    \end{bmatrix}
\end{aligned}
\end{equation}

Let $\hat{L}_{pp} = L_{pp} - h_p^T MH^{-1} h_p$ and $\hat{v}_1 = v_1 - h_p^T H^{-1}v_3$. Note that for equality constraint optimization problem, we get $\hat{L}_{pp} = L_{pp} $ and $\hat{v}_1 = v_1 $. We get the remaining vectors of $w$ as follows

\begin{equation}\label{eqn:updates}
\begin{aligned}
    w_2 & = \left[ g_p^T \hat{L}_{pp}^{-1}g_p \right]^{-1} [- v_2 + g_p \hat{L}_{pp}^{-1} \hat{v}_1] \\
    w_1 & = \hat{L}_{pp}^{-1} \left[ \hat{v}_1 - g_p^T w_2 \right] 
\end{aligned}
\end{equation}

Finally, we return the sensitivity vector $w$

\begin{equation}
\begin{aligned}
    \begin{bmatrix}
        \frac{dp ^*}{d \phi} \\[4pt]
        \frac{d\lambda ^*}{d\phi} \\[4pt]
        \frac{d \mu^*}{d \phi}\\[4pt]
    \end{bmatrix} v & = 
    \begin{bmatrix}
        w_1 \\[3pt]
        w_2 \\[3pt]
        w_3 \\[3pt]
    \end{bmatrix}
\end{aligned}
\end{equation}

\subsection{Derivative of outer optimization problem}

We consider the Lagrangian of the inner optimization problem as the objective of the outer optimization problem. We also assume that at the optimal solution of the inner optimization problem, none of the inequality constraints are active, i.e. $h(p^* | \phi) \neq 0 $ and therefore $ \mu ^* = 0$. These assumptions make the KKT point regular \citep{gros2020numerical} and simplify the computation of the gradient and Hessian of the outer objective with respect to $\phi$, as shown in Equation ~\ref{eqn:outer}.

\begin{equation}\label{eqn:outer}
\begin{aligned}
    \text{Outer Objective} & = L(p^*(\phi), \phi) \\
    \text{Gradient} & = \cancelto{0}{\frac{\partial L}{\partial p^*}} \frac{dp^*}{d\phi} + \frac{\partial L}{\partial \phi} \\
    \text{Hessian} & = \left(\frac{dp^*}{d\phi} \right)^T \frac{\partial ^2 L}{\partial {p^*}^2} \left( \frac{dp^*}{d\phi} \right) + \cancelto{0}{\frac{\partial L}{\partial p^*}} \frac{d^2p^*}{d\phi^2} + \frac{\partial ^2 L}{\partial \phi^2}
\end{aligned}
\end{equation}

Since we only require the gradient of $p^*$ with respect to $\phi$, its computation can be accelerated by storing the decomposition of the $\hat{L}_{pp}$ and $g_p \hat{L}_{pp} g_p$ during the forward pass, and reusing them when computing derivatives either using forward-mode (equation \ref{eqn:updates}) or reverse-mode automatic differentiation. However, storing this decomposition compromises the accuracy of any higher-order derivatives of $p^*$ with respect to $\phi$. Fortunately, this is acceptable in our case, as we do not need any higher-order derivatives as shown in Equation ~\ref{eqn:outer}. An additional advantage of reusing this decomposition, particularly in JAX, is that it makes the forward-mode equations \ref{eqn:updates} linear in the input tangent space \citep{frostig2021decomposingreversemodeautomaticdifferentiation}. As a result, a custom forward rule is sufficient for both forward- and reverse-mode automatic differentiation. This enables faster Hessian computation using the forward-over-reverse approach \citep{griewank2008evaluating}, compared to reverse-over-reverse mode in case when the decomposition is not reused. 

\subsection{Sensitivities of ordinary differential equations} \label{sec:senode}
Computing derivative of $p^*$ with respect to $\phi$ also requires computing sensitivities across the differential equation solver. Using the forward-mode optimize-then-discretize \cite{kidger2021on} differentiation approach gives

\begin{equation}
\begin{aligned}
    \frac{dX}{dt} & = \Theta(X, p^*(\phi), \phi) \\
    \frac{d}{d\phi}\frac{dX}{dt} & = \frac{d}{d\phi} \Theta(X, p^*(\phi), \phi)  = \frac{\partial \Theta}{\partial X} \frac{dX}{d\phi} + \frac{\partial \Theta}{\partial p^*} \frac{dp^*}{d\phi} + \frac{\partial \Theta}{\partial \phi} \\
    \frac{dS}{dt} & = \frac{\partial \Theta}{\partial X} S + \frac{\partial \Theta}{\partial p^*} \frac{dp^*}{d\phi} + \frac{\partial \Theta}{\partial \phi} 
\end{aligned}
\end{equation}

However, using interpolation makes the sensitivity calculations cheaper and relatively more stable by preventing sensitivities over unstable trajectories.  

\begin{equation}
\begin{aligned}
    \frac{dX}{dt} & = \Theta(\Psi(X)(t), p^*(\phi), \phi) \\
    \frac{d}{d\phi}\frac{dX}{dt} & = \frac{d}{d\phi} \Theta(\Psi(X)(t), p^*(\phi), \phi)  = \frac{\partial \Theta}{\partial p^*} \frac{dp^*}{d\phi} + \frac{\partial \Theta}{\partial \phi} \\
    \frac{dS}{dt} & = \frac{\partial \Theta}{\partial p^*} \frac{dp^*}{d\phi} + \frac{\partial \Theta}{\partial \phi} 
\end{aligned}
\end{equation}

For parameter estimation in delayed differential equations, as discussed in the next section, the sensitivities, using interpolation, are calculated as follows 

\begin{equation}
\begin{aligned}
    \frac{dX}{dt} & = \Theta(\Psi(X)(\tau), p^*(\phi), \phi) \\
    \frac{d}{d\phi}\frac{dX}{dt} & = \frac{d}{d\phi} \Theta(\Psi(X)(t), p^*(\phi), \phi) \\
    \frac{dS}{dt} & = \frac{\partial \Theta}{\partial \Psi}\frac{d\Psi}{d\tau}\frac{d\tau}{d\phi} + \frac{\partial \Theta}{\partial p^*} \frac{dp^*}{d\phi} + \frac{\partial \Theta}{\partial \phi} \\
\end{aligned}
\end{equation}

where $\tau = t - \phi$ and $\frac{d\Psi}{d\tau}$ is the derivative of the interpolation with respect to time. These results hold for reverse-mode and discretize-then-optimize differentiation approaches.

\section{Experiments}

We demonstrate the efficacy of our method on two different problem classes using benchmark problems \cite{calver2019parameter}. These problems are highly nonlinear, coupled, and sensitive to parameter variations, all of which make the optimization problem challenging and prone to convergence issues or poor local minima. First, we consider parameter estimation, where the goal is to estimate the parameters of a system of ordinary differential equations using state measurements. In this setting, we assume that the structure of the function $\Theta (X, p, \phi)$ is known, while the parameters $p$ and $\phi$ are unknown. Second, we consider model discovery, in which both the parameters and the function $\Theta (X, p, \phi)$ are unknown. To make this problem tractable, we assume that the underlying function can be represented as a sparse combination of basis functions selected from a large candidate library, akin to SINDy and recent strong implementations of this method \cite{prabhu2025derivative}. Consequently, the problem reduces to selecting the appropriate basis functions from this library and subsequently estimating the associated coefficients (parameters). Finally, we demonstrate how our method can be extended to parameter estimation of delayed differential equations, stiff differential equations, and partially observed differential equations.

The code is made available online at \url{https://github.com/siddharth-prabhu/BiLevelParameterEstimation}. We use SciPy \citep{2020SciPy-NMeth} for interpolations, IPOPT \citep{wachter2006implementation} for nonlinear optimization, JAX \citep{jax2018github} for derivative calculations, and Pyomo \citep{hart2011pyomo, Nicholson2018} for implementing orthogonal collocation. The examples considered in this study include

\begin{enumerate}
    \item Oscillatory dynamics of \textbf{Calcium ion} (parameter estimation, ordinary differential equation): This is a four-state ODE describing the dynamics of calcium ion in eukaryotic cells \cite{kummer2000switching} comprising linear, bilinear, and Michaelis-Menton type rate terms. The detailed model, parameters, and data generation strategy is given in Appendix \ref{appendix:calcium} 
    \item Nonlinear biochemical model studied by \textbf{Mendes} \citep{moles2003parameter}. (parameter estimation, ordinary differential equation):  This 8-state 36 (15 linear and 21 nonlinear) parameter ODE model describes the variation of metabolite concentrations with time. The model equation, parameters, and the data sampling strategies are in Appendix \ref{appendix:mendes}. This comprises linear and nonlinear terms in state variables. 
    \item \textbf{Ethanol fermentation} (parameter estimation, ordinary differential equation): This system describes the growth of microorganisms, the consumption of glucose and the formation of the products in a batch fermentation process \citep{wang2001hybrid}. This system has Michaelis-Menton kinetics with linear/bilinear terms in state variables as well as  quadratic terms in the denominator. The model equations, parameters, and the data sampling strategies are in Appendix \ref{appendix:ferment}. 
    \item \textbf{Kermack-McKendrick} disease-spreading models (parameter estimation, delayed differential equation): This is an extension of infectious disease spreading (susceptible, infected, recovered models) comprising linear and bilinear terms, while also including time delays \citep{doi:10.1137/120889733}. The model equations, parameters, and the data sampling strategies are in Appendix \ref{appendix:km}. 
    \item \textbf{Belousov reaction} (parameter estimation, stiff ordinary differential equation): This is a classic example of a chemical oscillator in a continuously stirred batch reactor, where the concentrations of certain intermediates cycle repeatedly between high and low values rather than simply decaying to equilibrium \cite{gray2002analysis}. Stiff systems arise in the modeling of multi–timescale dynamics, where at least one state evolves much more slowly or much more rapidly than the others. As a consequence, the corresponding parameters often span several orders of magnitude. Parameter estimation for stiff ordinary differential equations is particularly challenging due to the high computational cost of solving stiff ODEs, as well as instability and ill-conditioning in the associated optimization problem \cite{doi:10.1137/1021001, PhysRevE.90.023303, vilela2009identification}. The proposed approach allows stiff systems to be handled without requiring stiff ODE solvers during optimization. The model equations, parameters, and the data sampling strategies are in Appendix \ref{appendix:br}. 
    \item \textbf{Continuous stirred tank reactor} (parameter estimation, partially observed ordinary differential equation): This system exhibits strong nonlinearity and, depending on the operating conditions, can display multiple steady states, limit cycles, or even chaotic behavior, making it a demanding test case for identification algorithms. While our approach builds on interpolation over measured states, it is still possible to perform parameter estimation for differential equations with partially observed states, such as latent or difficult to measure states. Note that, in this example, we assume that the parameters are identifiable \cite{kravaris2013advances}. There are two main ways to address this problem. The first approach assumes trajectories of the unobserved states at selected time points as decision variables in the optimization problem. A cubic-spline interpolation is then fitted to these values (which corresponds to solving a linear system of equations followed by polynomial evaluation at a given time argument, and is therefore fully differentiable), after which bi-level optimization is performed. In this formulation, the nonlinear parameters include both the nonlinear model parameters and the values of the unobserved states at the chosen time points. Although feasible, this strategy is computationally expensive because gradients and Hessians must be computed through the interpolation procedure. Furthermore, the interpolation method must be compatible with the automatic-differentiation library. An implementation of cubic-spline interpolation in JAX can be found here \url{https://github.com/siddharth-prabhu/BiLevelParameterEstimation/blob/main/utils.py}. The second approach uses sequential optimization (single shooting) for the unobserved states, combined with the proposed method for the observed states. Despite not exploiting interpolation during sensitivity calculations, as discussed in Section \ref{sec:senode}, this approach may still be computationally cheaper. The model equations, parameters, and the data sampling strategies are given in Appendix \ref{appendix:cstr}. 
    \item \textbf{Esterification of carboxylic acid} (model discovery, ordinary differential equation) : This example is considered because it poses a significant challenge for model discovery methods, because of the large number of states, the complexity of the dynamics, and the sparsity of the governing equations. The model equations, parameters, and the data sampling strategies are given in Appendix \ref{appendix:carb}. In this problem, we seek to estimate the kinetic parameters (reaction rate constants and activation energies) as well as the functional form of the reaction rate. We assume that this function is an additive combination of a subset of basis functions drawn from a library of all possible polynomial combinations of the states \citep{brunton2016discovering}. Thus, the problem reduces to selecting the correct terms for each reaction and estimating their corresponding coefficients. We use a thresholding-based algorithm to eliminate these spurious terms \citep{brunton2016discovering}. We consider two possible convergence criteria ($T$). The first is tolerance-based, where an optimization problem is initially solved with a lower tolerance, and the tolerance is gradually increased in each thresholding cycle. The second is iteration-based, where the initial optimization problem is solved with fewer iterations, while the later problem is solved until convergence. Once the convergence criterion is met, the parameters smaller than the thresholding parameter ($\epsilon$) are eliminated, and the optimization problem is repeated with the remaining coefficients. We also consider two possible terminating conditions 1) all the parameters are eliminated, in which case the thresholding parameter is too large to consider all the coefficients and no solution is obtained 2) there are no more coefficients to be eliminated, in which case the optimal solution with the remaining coefficients has been found. The steps are outlined in algorithm \ref{algo:stlsq}. 
    
    \begin{algorithm}[h!]
        \caption{Sequential Threshold Bi-Level Optimization}\label{algo:stlsq}
        \begin{algorithmic}
        \Require Measurement matrix $ \hat{X}$, ridge penalty $\lambda$, thresholding parameter $\epsilon$, convergence criterion $T$
        \State $ \psi _{guess}, \ p_{guess} \sim \mathcal{U}(\epsilon, \infty)$ \Comment{Initialize randomly}
        \While{not converged}
        \State $ \phi ^*, p ^* \gets \text{Optimize Equation \ref{eqn:library} using bi-level optimization with ridge penalty on $p$} $
        \State $ p ^* [\text{abs}(p ^*) \leq \epsilon] \gets  0 $ \Comment{Threshold small indices to zero}
        \State $p _{guess} \gets p ^* [\text{abs}(p ^*) > \epsilon] $ \Comment{Choose big coefficients}
        \State $T \gets $ Update convergence criterion
        \EndWhile
        \State \Return $ p^*, \phi ^*$ 
        \end{algorithmic}
    \end{algorithm}

\end{enumerate}

\section{Results}

We compare the performance of our proposed bi-level method with a sequential optimization (single-shooting) and orthogonal collocation. In the single-shooting method, we solve a nonlinear optimization problem over all the parameters. The trajectory is obtained by integrating the differential equations using the current iterate of the parameters, and the parameters are then updated using gradients computed across this integration. For both methods, i.e. single-shooting and our proposed bi-level method, we use the explicit ODE solver Dormand–Prince (RKDP) \cite{dormand1980family} for integration and an optimize-then-discretize approach to compute gradients across the integration \cite{kidger2021on}. Orthogonal collocation, on the other hand, discretizes the trajectory into finite elements and approximates the state variables within each element as a linear combination of polynomial basis functions, evaluated at specially chosen collocation points. This eliminates the need to explicitly integrate the ODEs, reducing the dynamic optimization problem to a system of algebraic equations solved within a single optimization framework. Further details on the optimization problem formulation are provided in Appendix \ref{appendix:opti}.

Table \ref{tab:results} reports the performance of the methods using several metrics. First, Status indicates whether the optimization problem converged to an optimal solution or failed to converge. If convergence is achieved, we report the optimal solution, otherwise, we report the solution obtained thus far. A dash ($-$) is used to indicate that the method failed to find a feasible point. Second, $J$ denotes the mean-squared error between the simulated trajectories generated using the estimated parameters and the true trajectory. Third, Iterations reports the number of iterations required to obtain the reported solution. Finally, Wall Time gives the total runtime in seconds required for the algorithm to converge (or fail to converge), evaluated on a Linux-based system with 10 cores running at 2.3 GHz. Note that only a subset of the benchmark problems are used for direct comparison, while the remaining systems serve to demonstrate the flexibility of the proposed method in handling diverse problems arising in real-world applications.

\begin{figure}[!ht]
\centering
    \begin{subfigure}{0.19\textwidth}
      \includegraphics[width = \linewidth, trim = 280 200 280 200, clip]{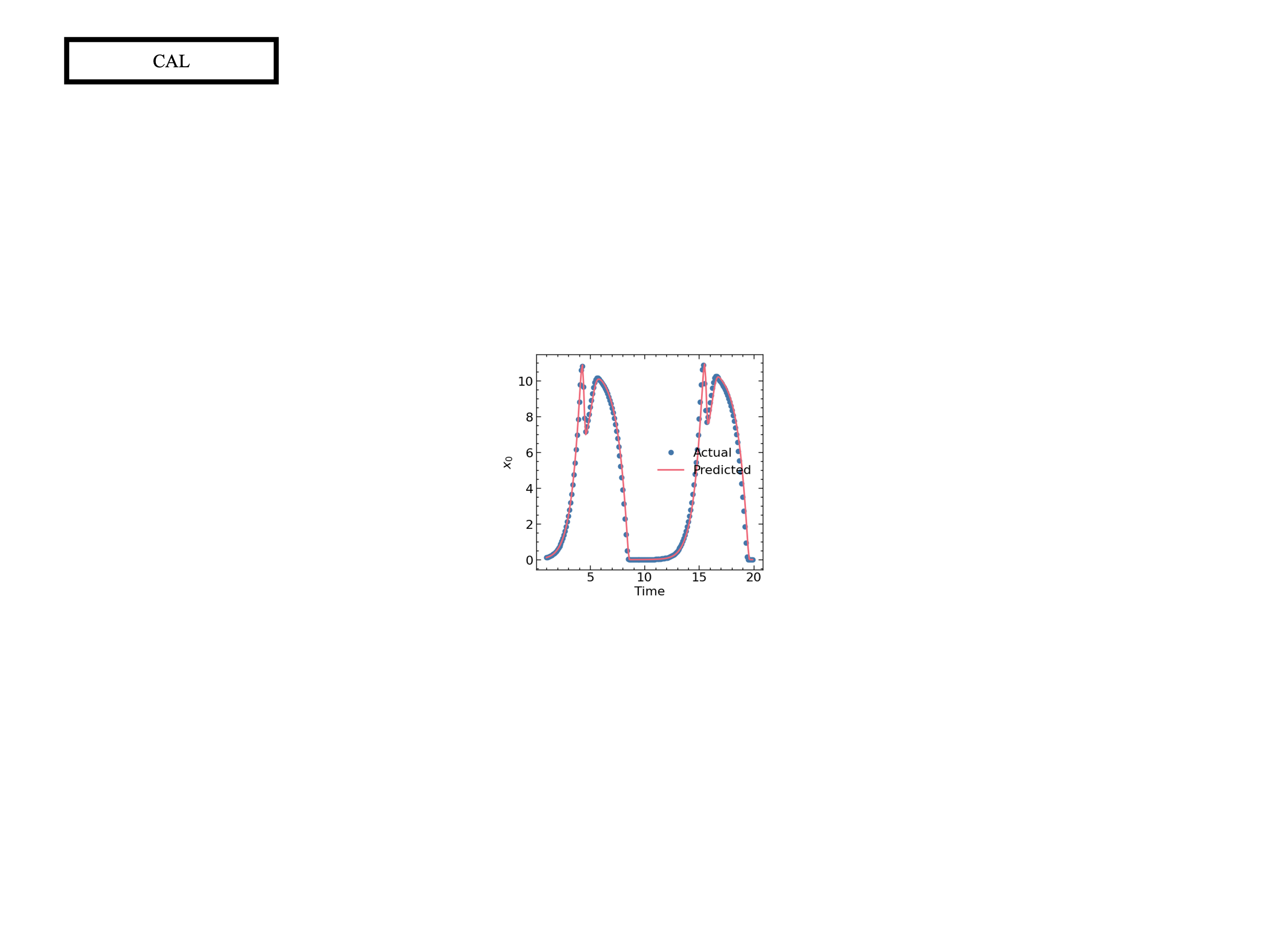}
      \caption{}
    \end{subfigure}
    \bigskip
    \centering
    \begin{subfigure}{0.19\textwidth}
      \includegraphics[width = \linewidth, trim = 280 200 280 200, clip]{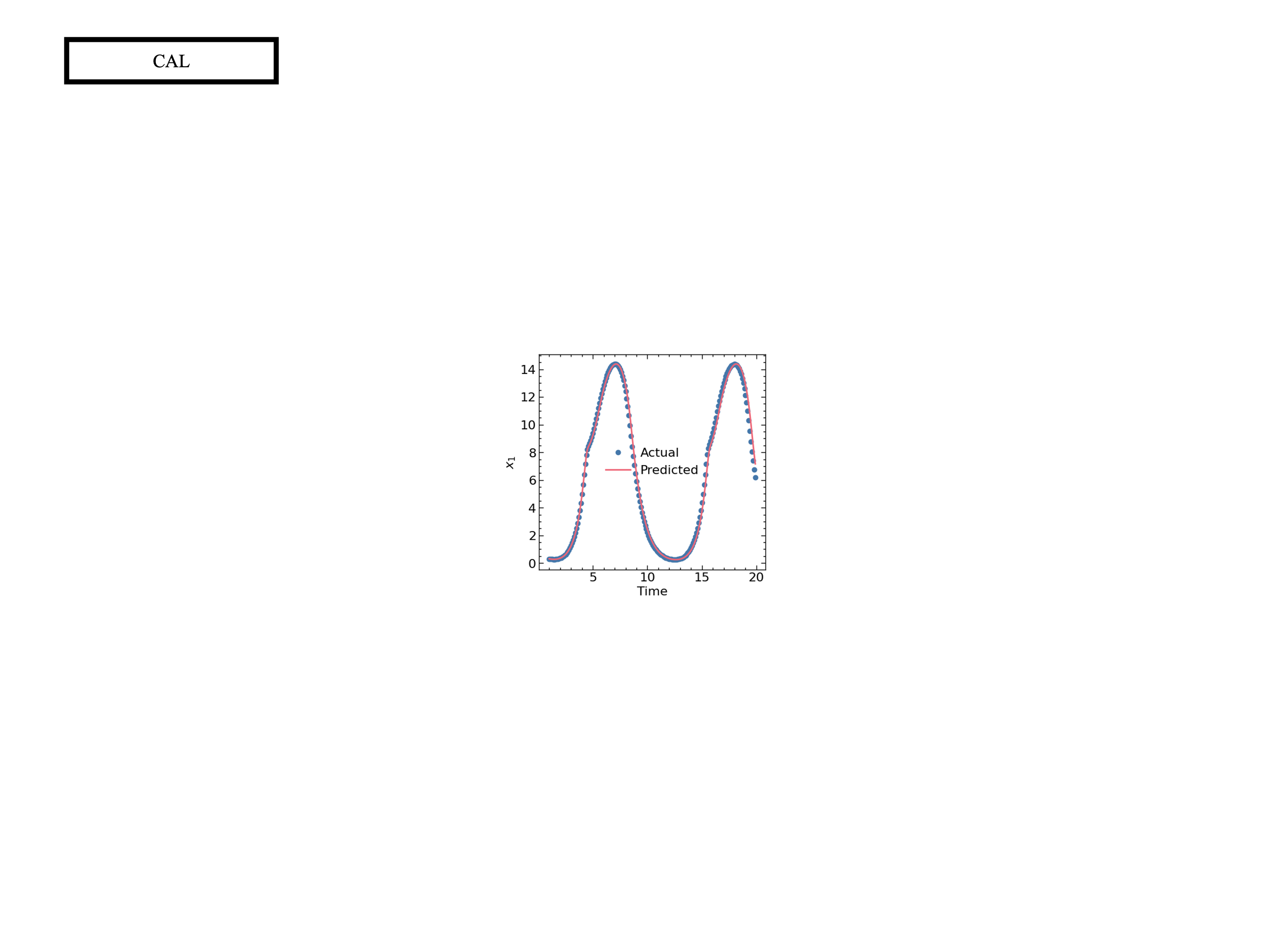}
      \caption{}
    \end{subfigure}
    \centering
    \begin{subfigure}{0.19\textwidth}
      \includegraphics[width = \linewidth, trim = 280 200 280 200, clip]{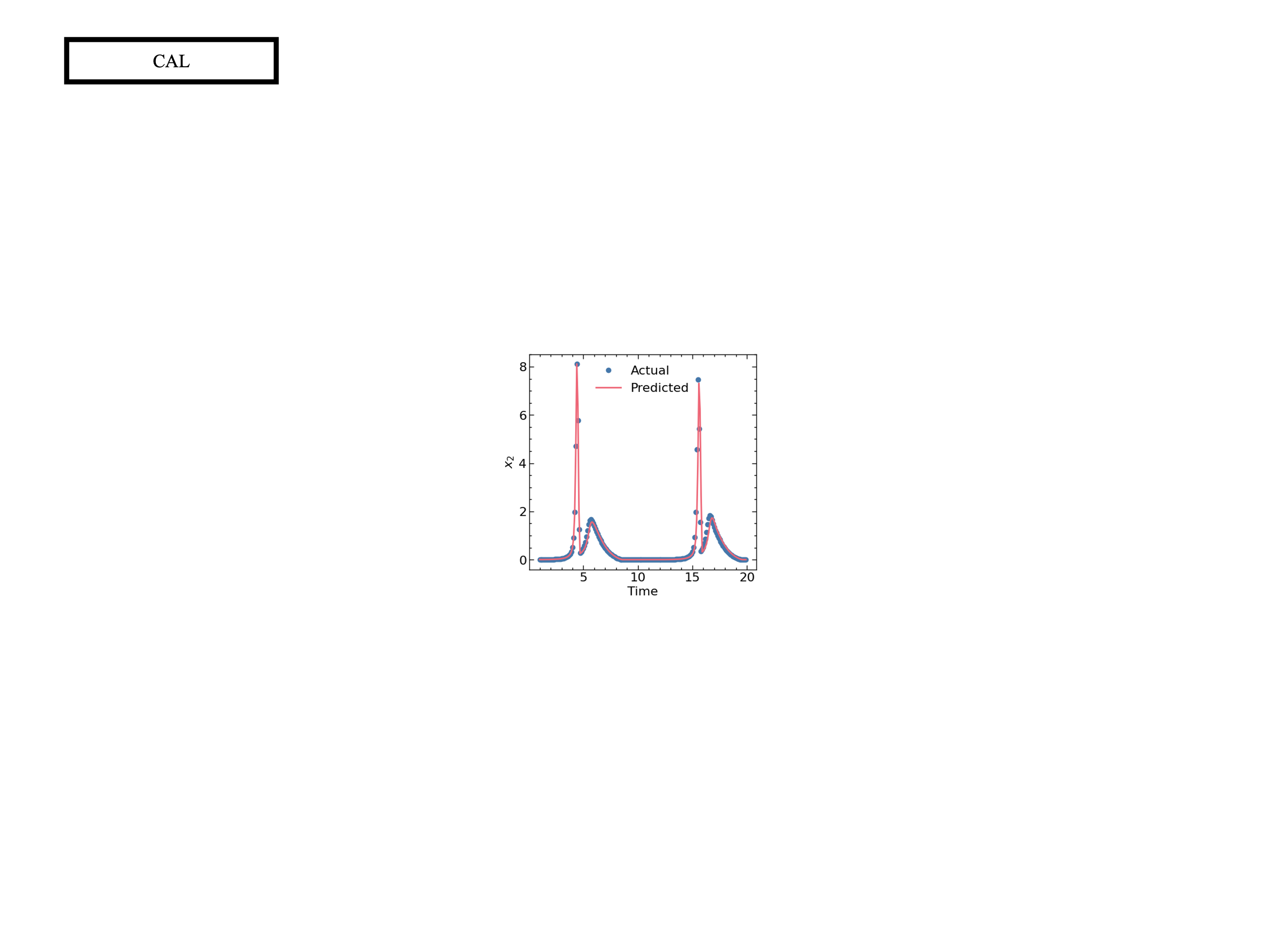}
      \caption{}
    \end{subfigure}
    \centering
    \begin{subfigure}{0.19\textwidth}
      \includegraphics[width = \linewidth, trim = 280 200 280 200, clip]{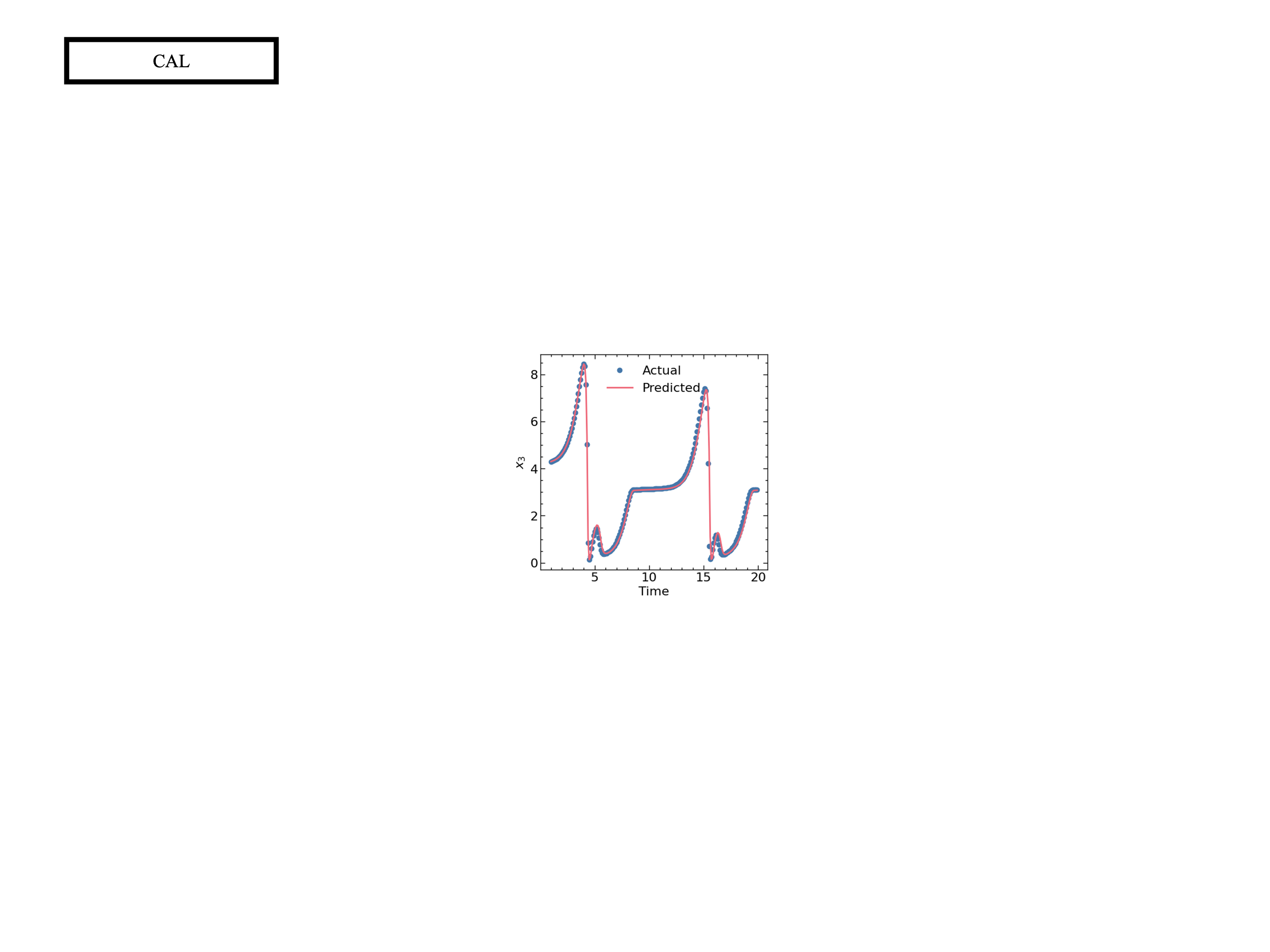}
      \caption{}
    \end{subfigure}
    \centering
    \begin{subfigure}{0.19\textwidth}
      \includegraphics[width = \linewidth, trim = 280 200 280 200, clip]{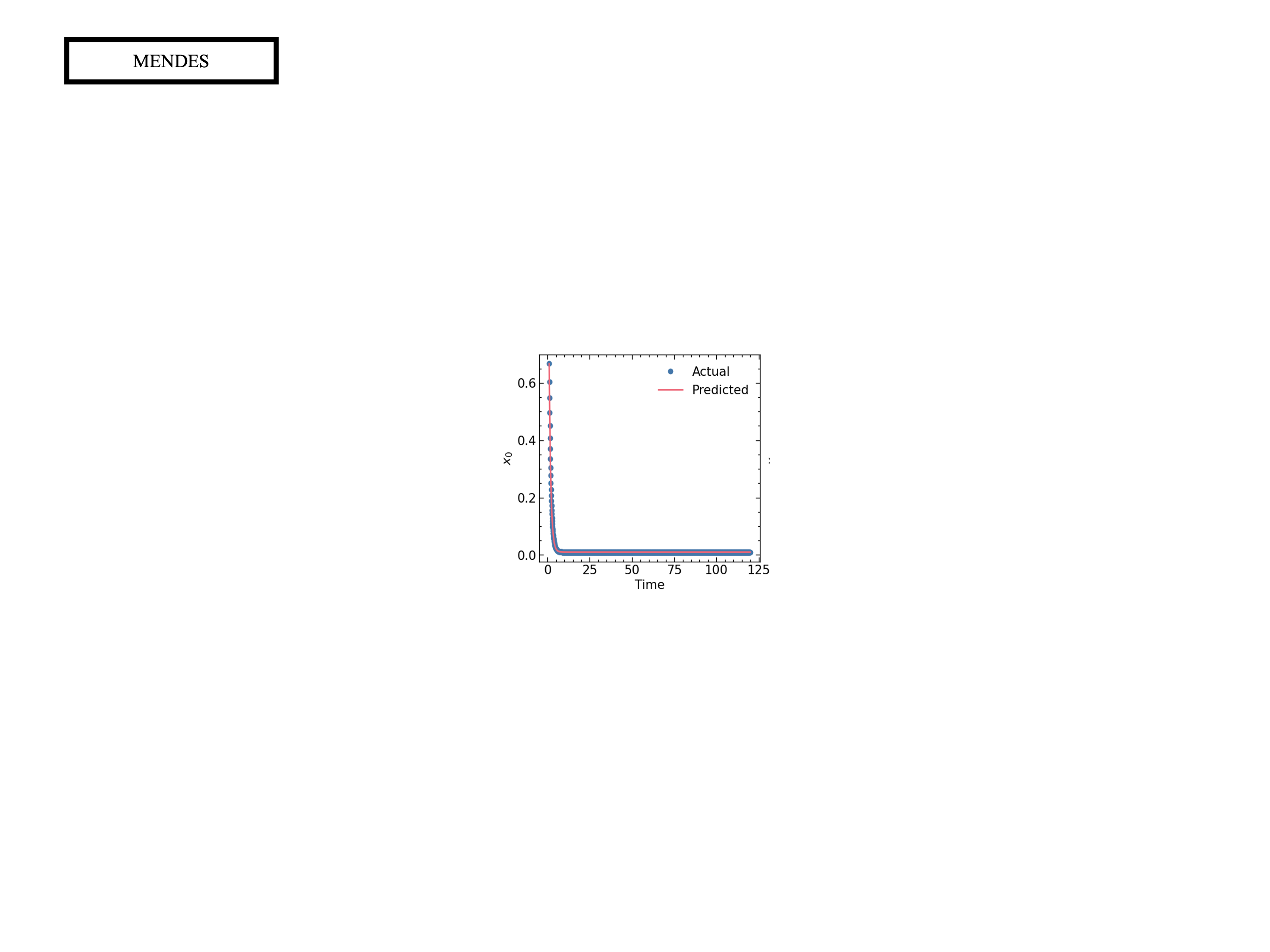}
      \caption{}
    \end{subfigure}
    \bigskip
    \centering
    \begin{subfigure}{0.19\textwidth}
      \includegraphics[width = \linewidth, trim = 280 200 280 200, clip]{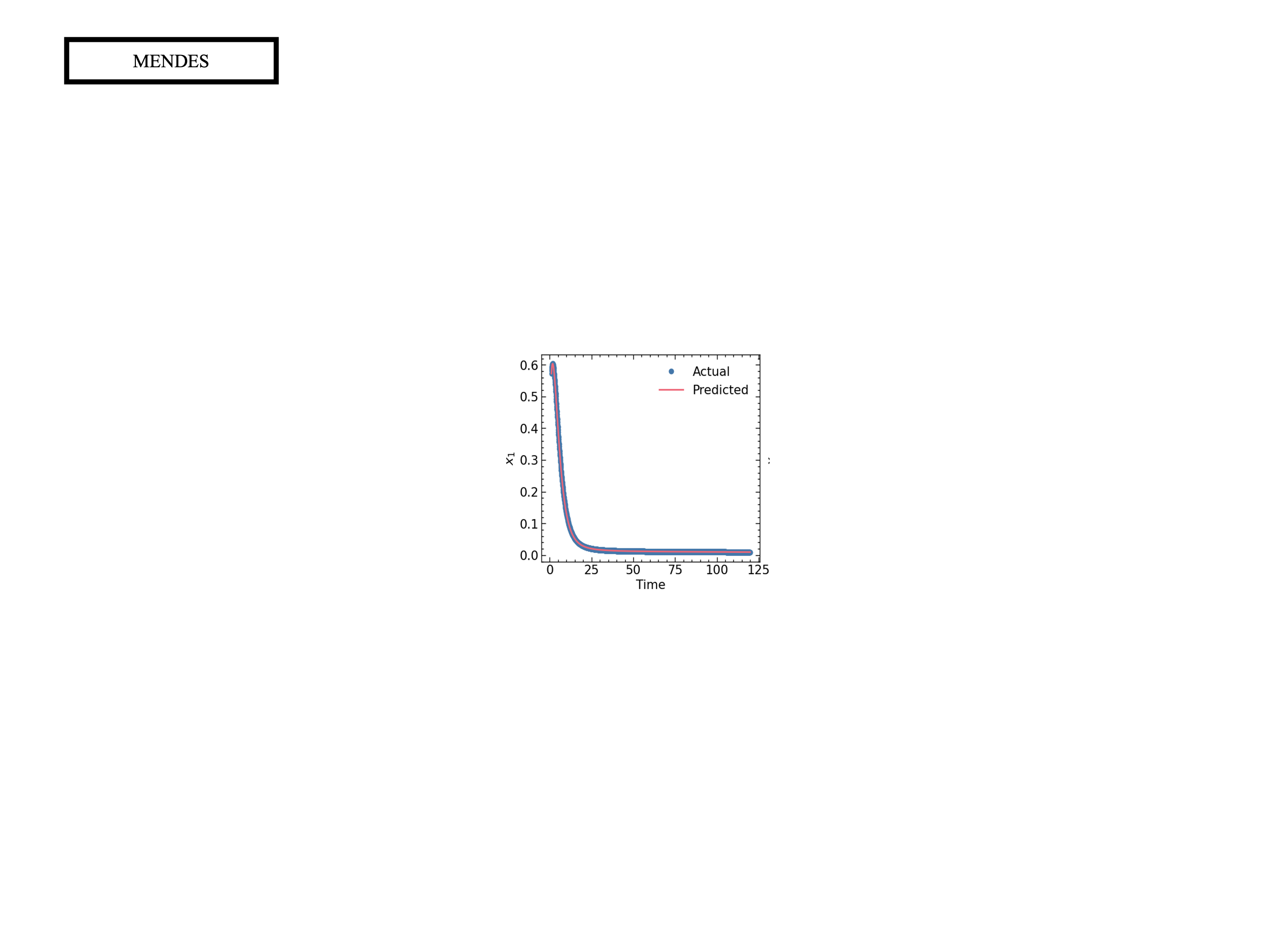}
      \caption{}
    \end{subfigure}
    \centering
    \begin{subfigure}{0.19\textwidth}
      \includegraphics[width = \linewidth, trim = 280 200 280 200, clip]{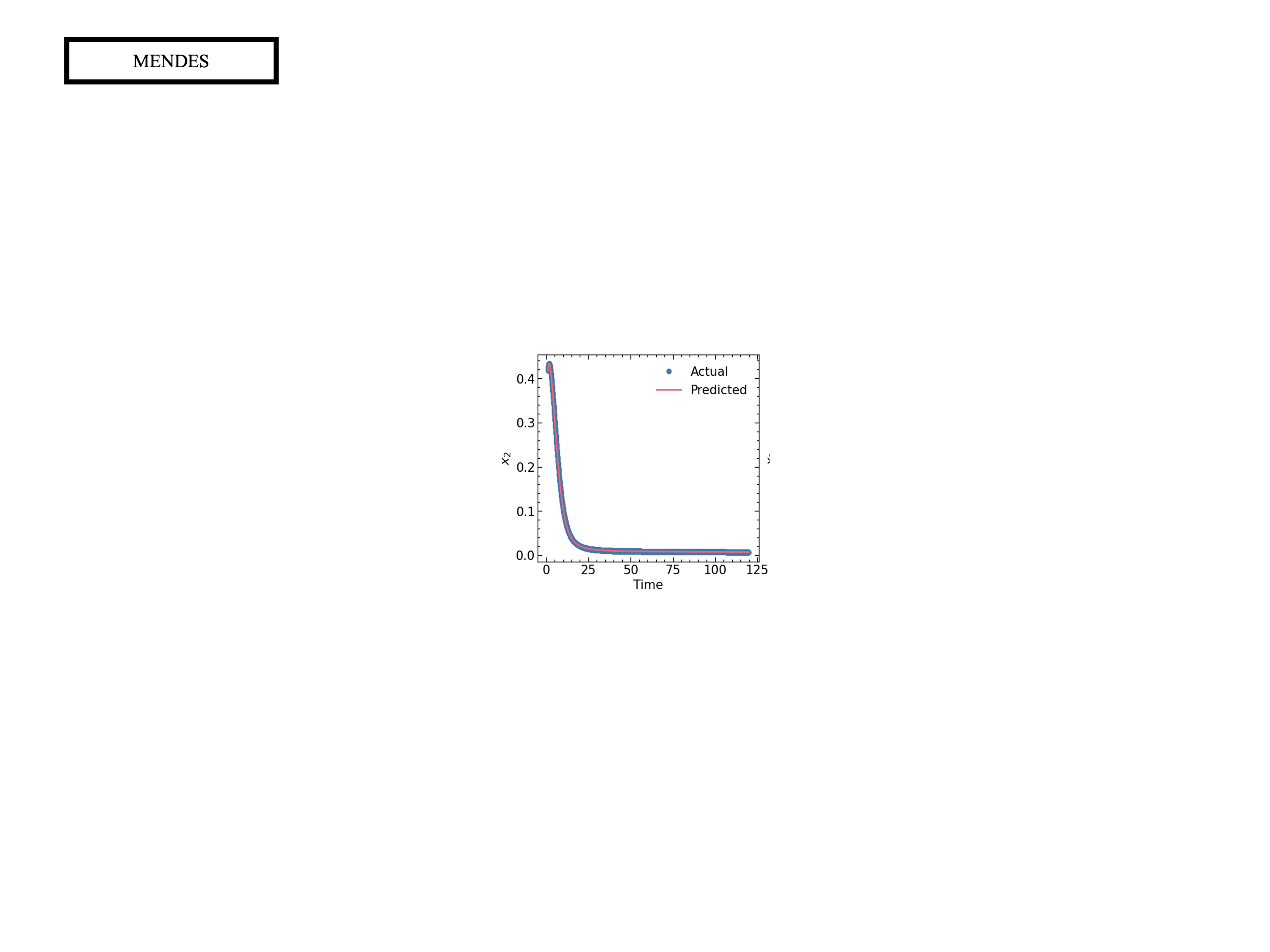}
      \caption{}
    \end{subfigure}
    \centering
    \begin{subfigure}{0.19\textwidth}
      \includegraphics[width = \linewidth, trim = 280 200 280 200, clip]{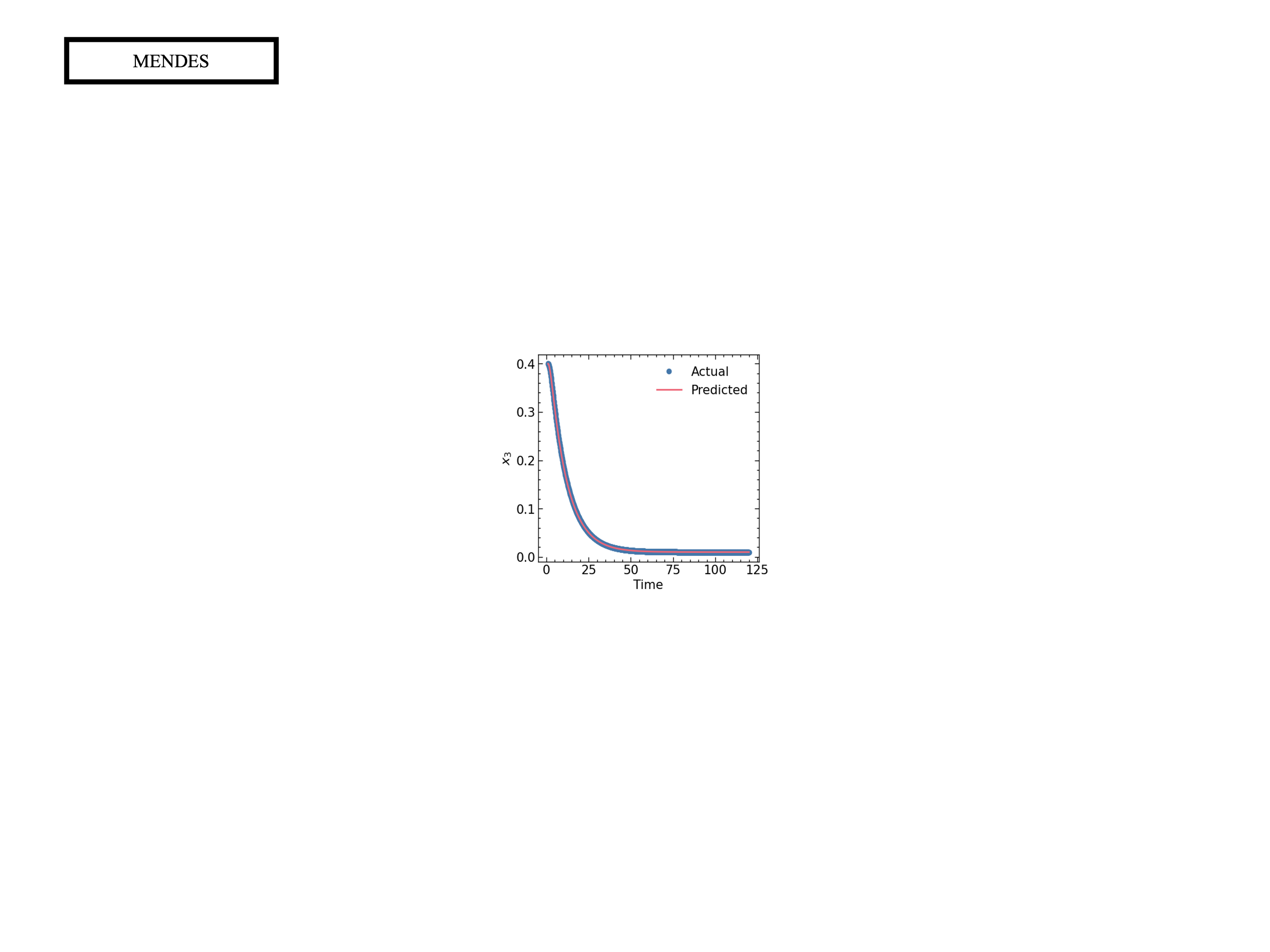}
      \caption{}
    \end{subfigure}
    \centering
    \begin{subfigure}{0.19\textwidth}
      \includegraphics[width = \linewidth, trim = 280 200 280 200, clip]{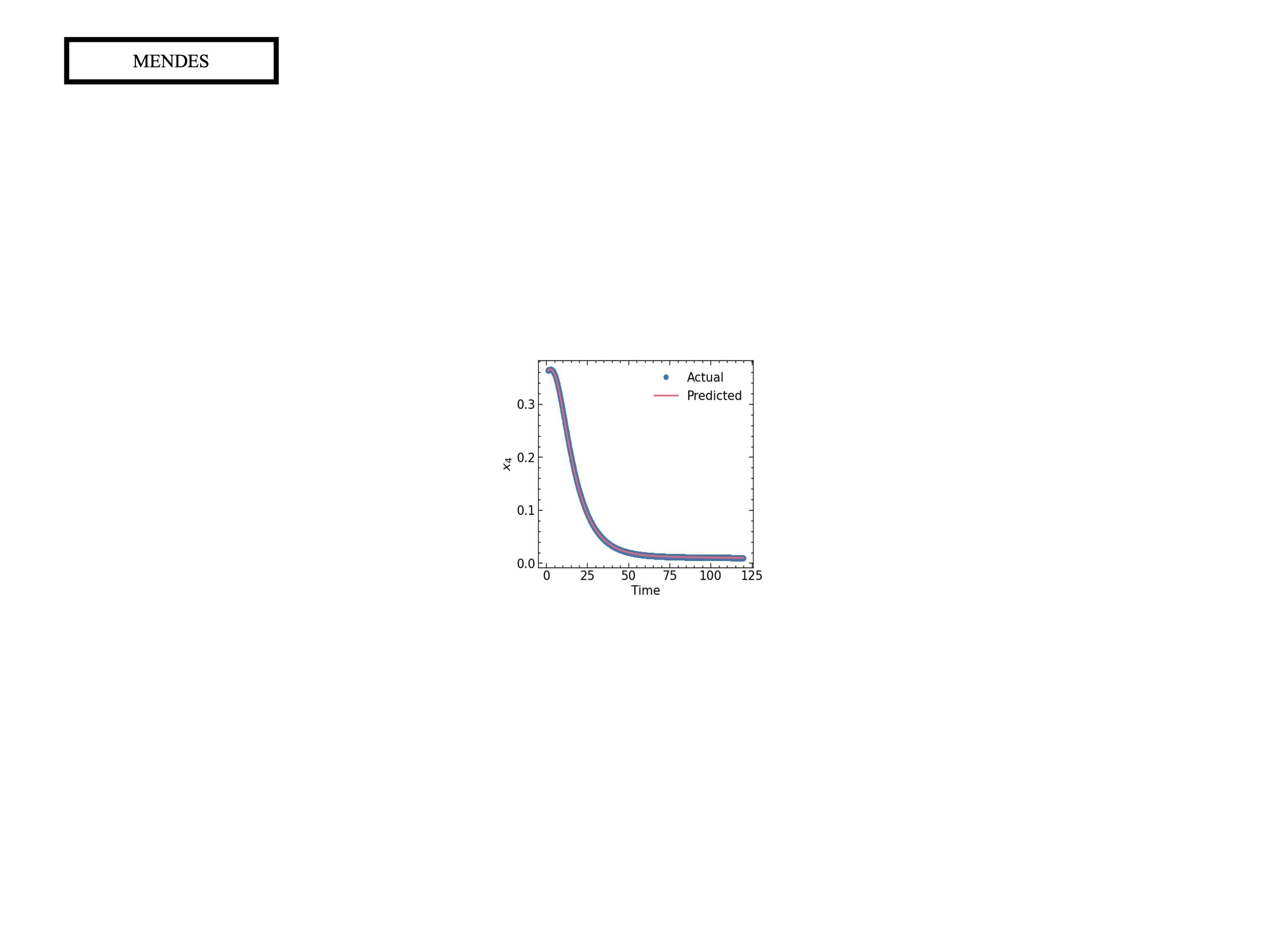}
      \caption{}
    \end{subfigure}
    \centering
    \begin{subfigure}{0.19\textwidth}
      \includegraphics[width = \linewidth, trim = 280 200 280 200, clip]{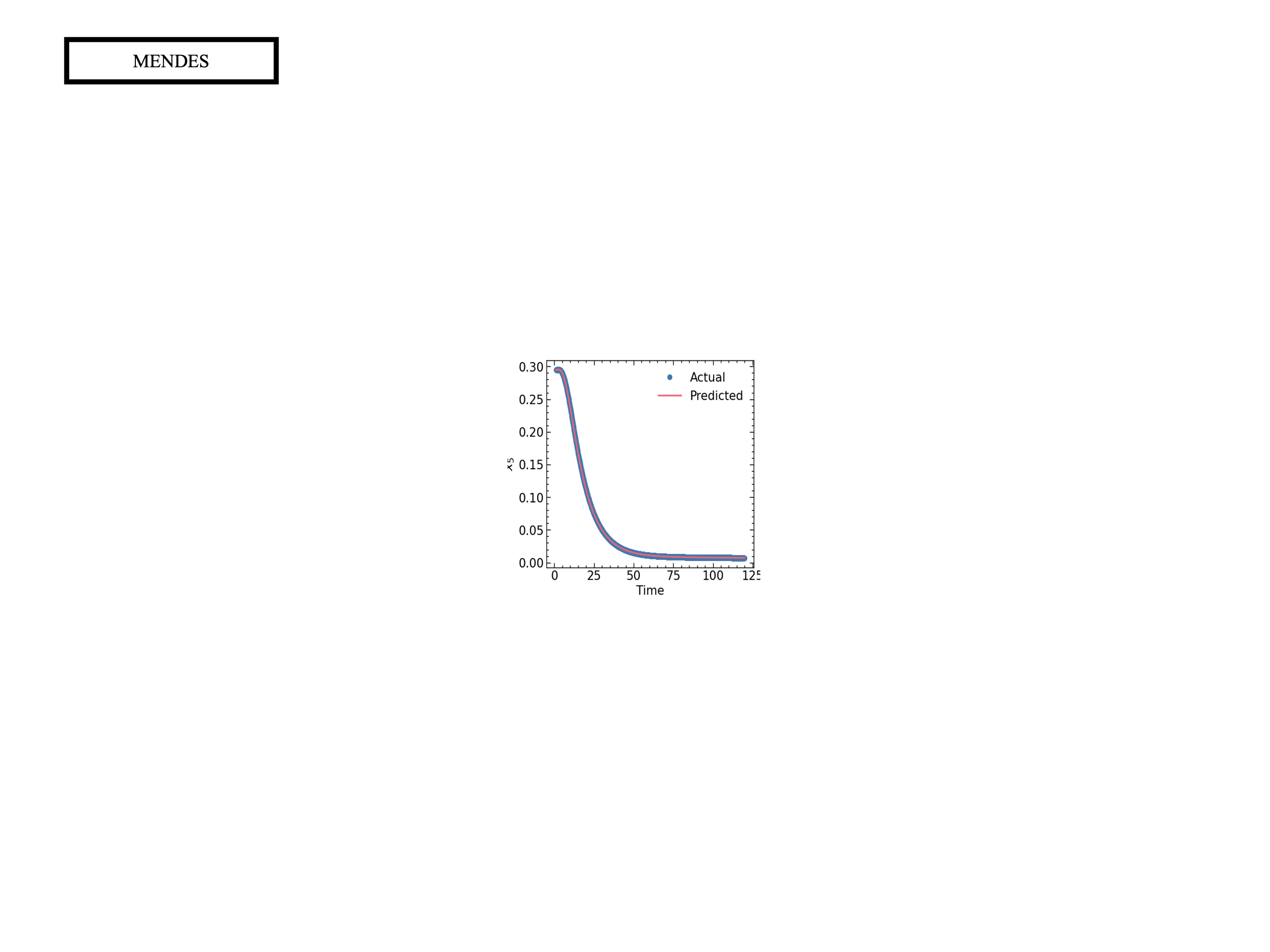}
      \caption{}
    \end{subfigure}
    \bigskip
    \centering
    \begin{subfigure}{0.19\textwidth}
      \includegraphics[width = \linewidth, trim = 280 200 280 200, clip]{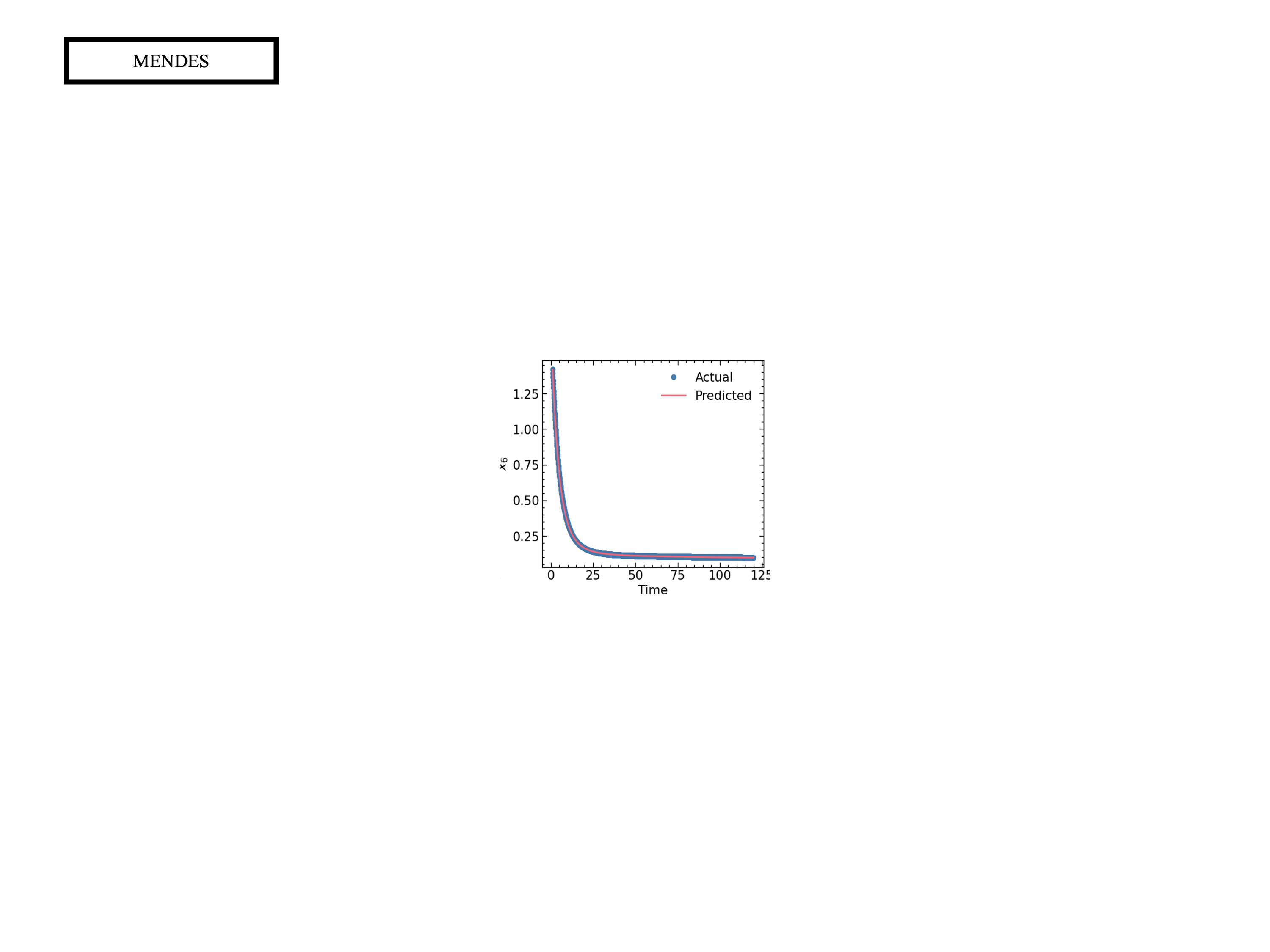}
      \caption{}
    \end{subfigure}
    \centering
    \begin{subfigure}{0.19\textwidth}
      \includegraphics[width = \linewidth, trim = 280 200 280 200, clip]{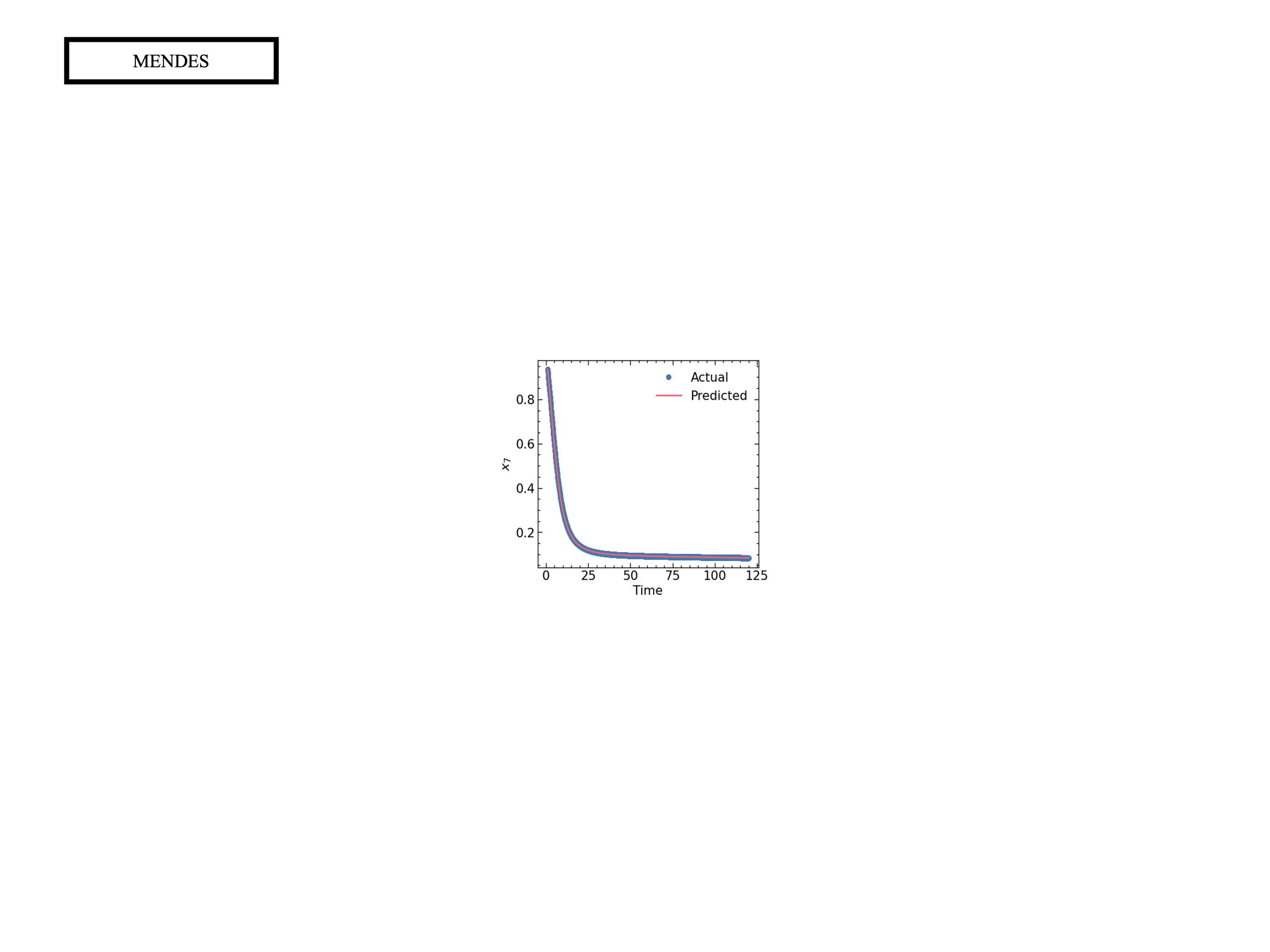}
      \caption{}
    \end{subfigure}
    \centering
    \begin{subfigure}{0.19\textwidth}
      \includegraphics[width = \linewidth, trim = 280 200 280 200, clip]{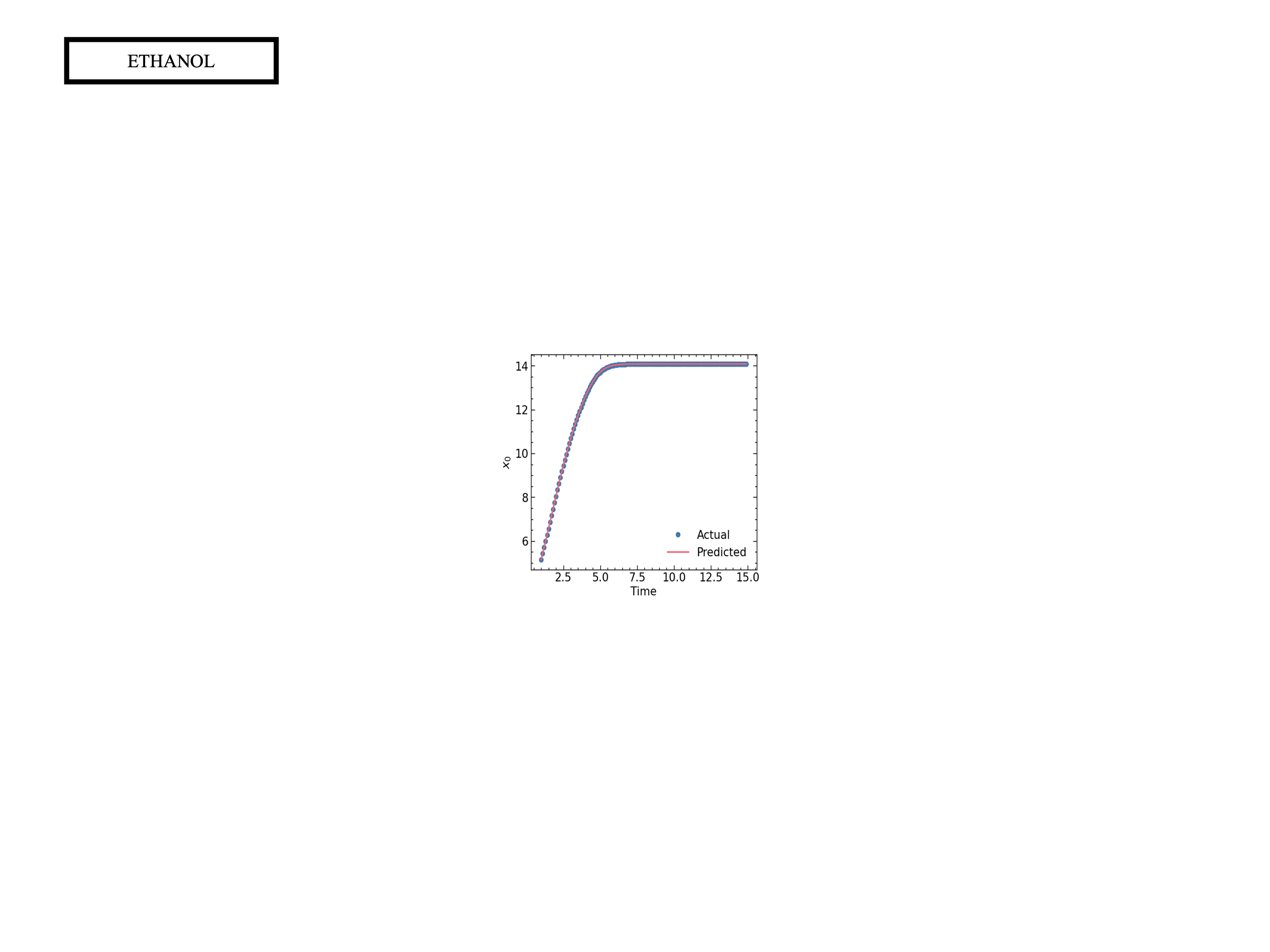}
      \caption{}
    \end{subfigure}
    \centering
    \begin{subfigure}{0.19\textwidth}
      \includegraphics[width = \linewidth, trim = 280 200 280 200, clip]{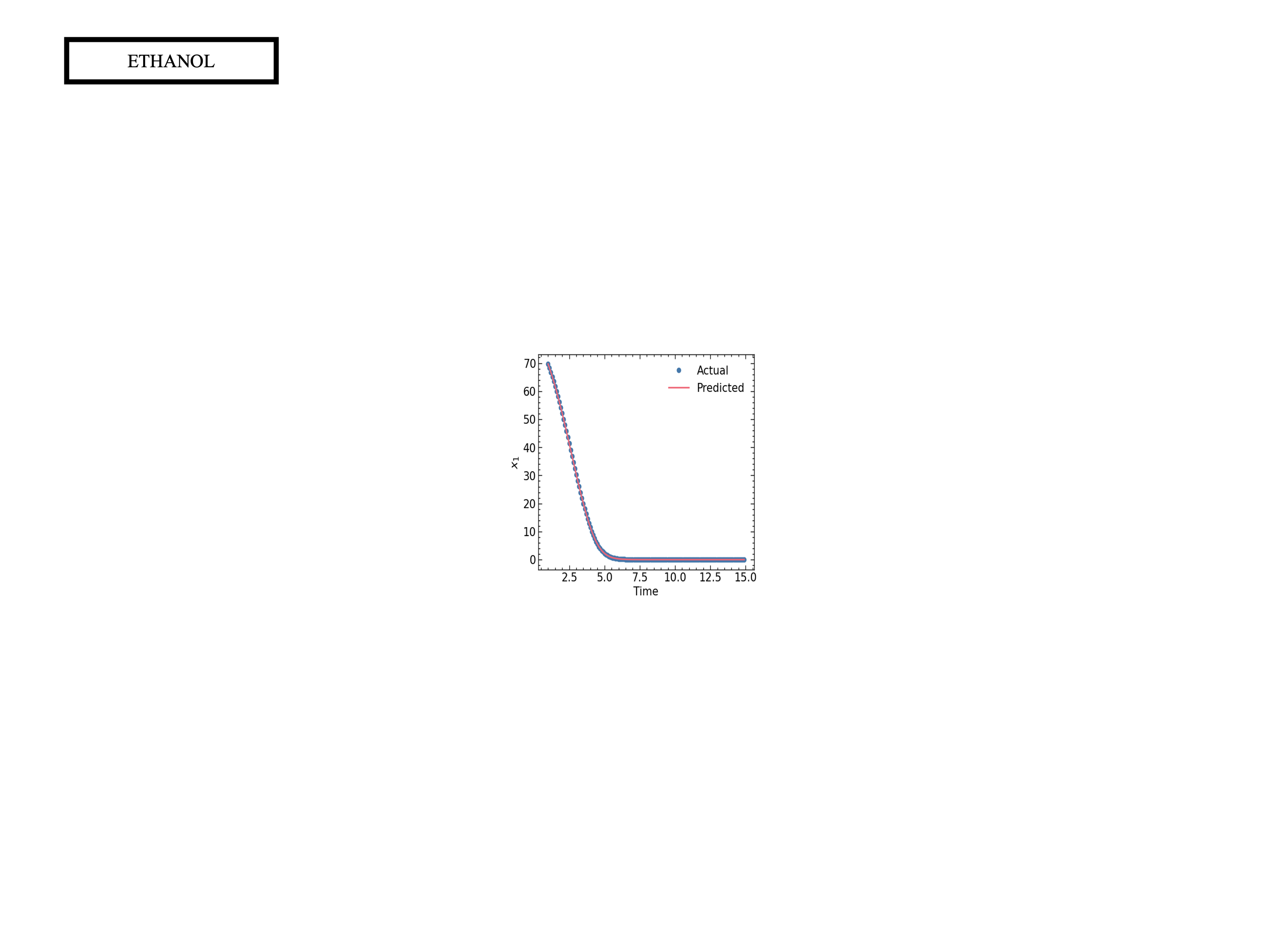}
      \caption{}
    \end{subfigure}
    \centering
    \begin{subfigure}{0.19\textwidth}
      \includegraphics[width = \linewidth, trim = 280 200 280 200, clip]{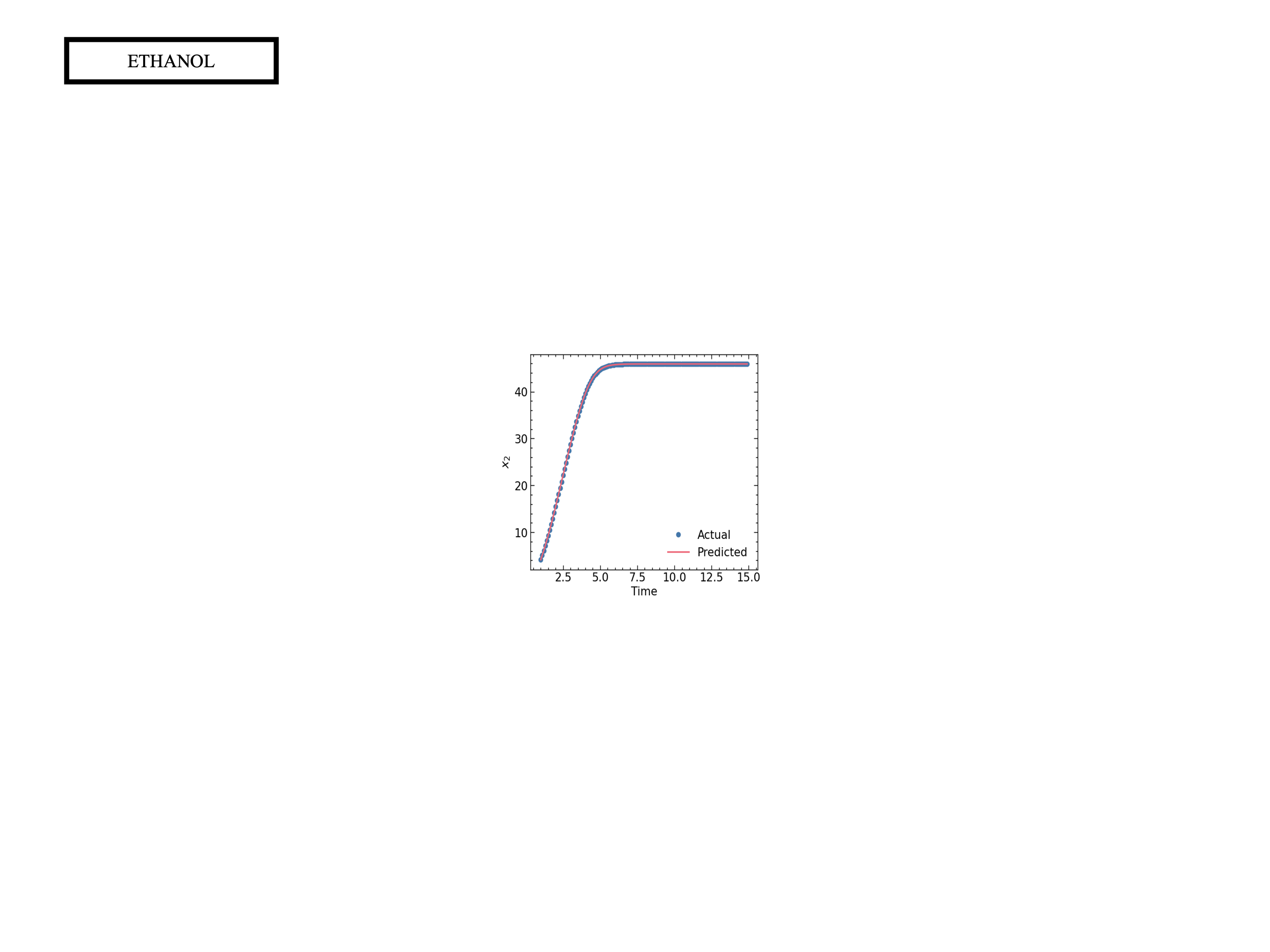}
      \caption{}
    \end{subfigure}
    \bigskip
    \centering
    \begin{subfigure}{0.19\textwidth}
      \includegraphics[width = \linewidth, trim = 280 200 280 200, clip]{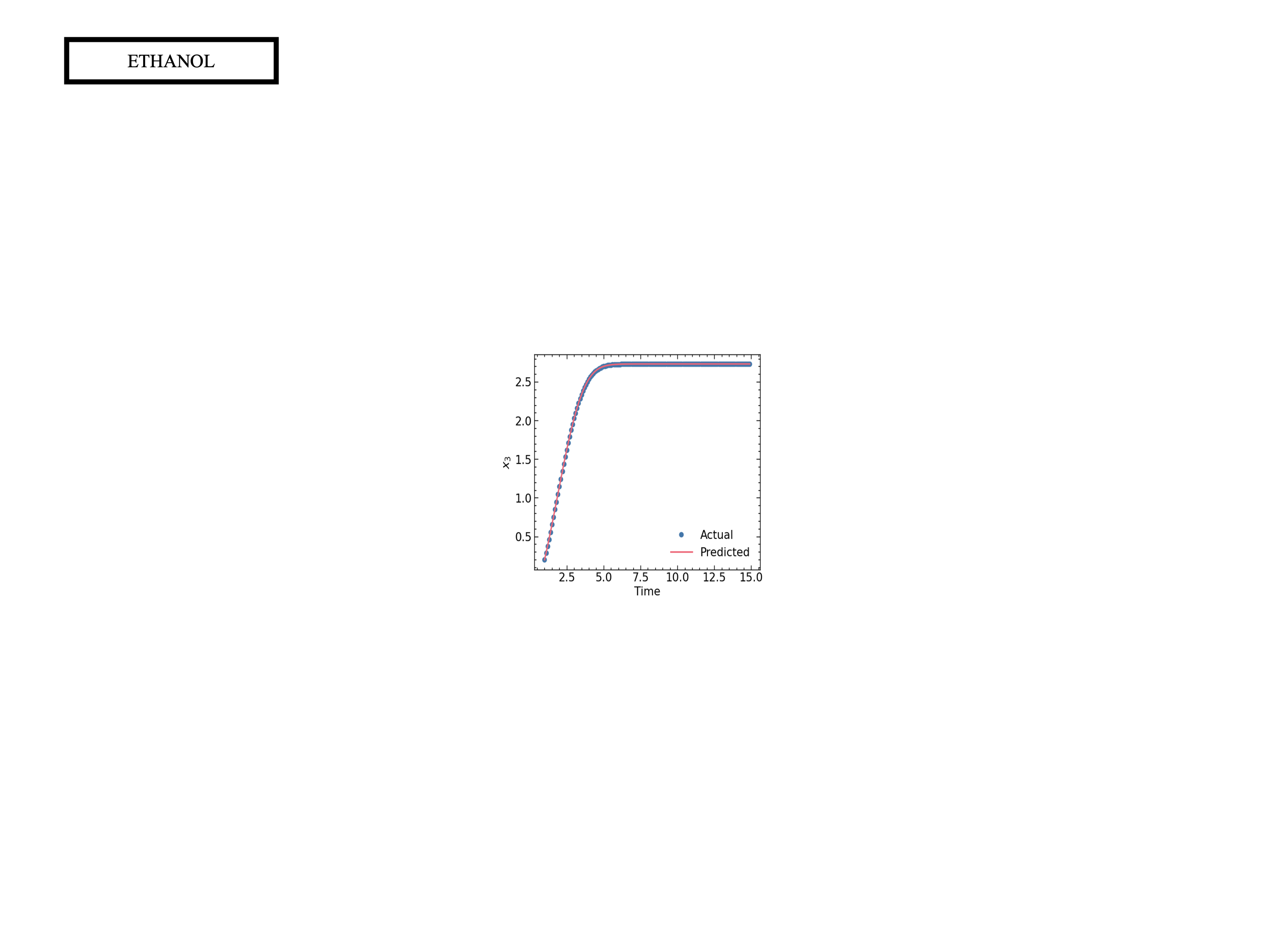}
      \caption{}
    \end{subfigure}
    \centering
    \begin{subfigure}{0.19\textwidth}
      \includegraphics[width = \linewidth, trim = 280 200 280 200, clip]{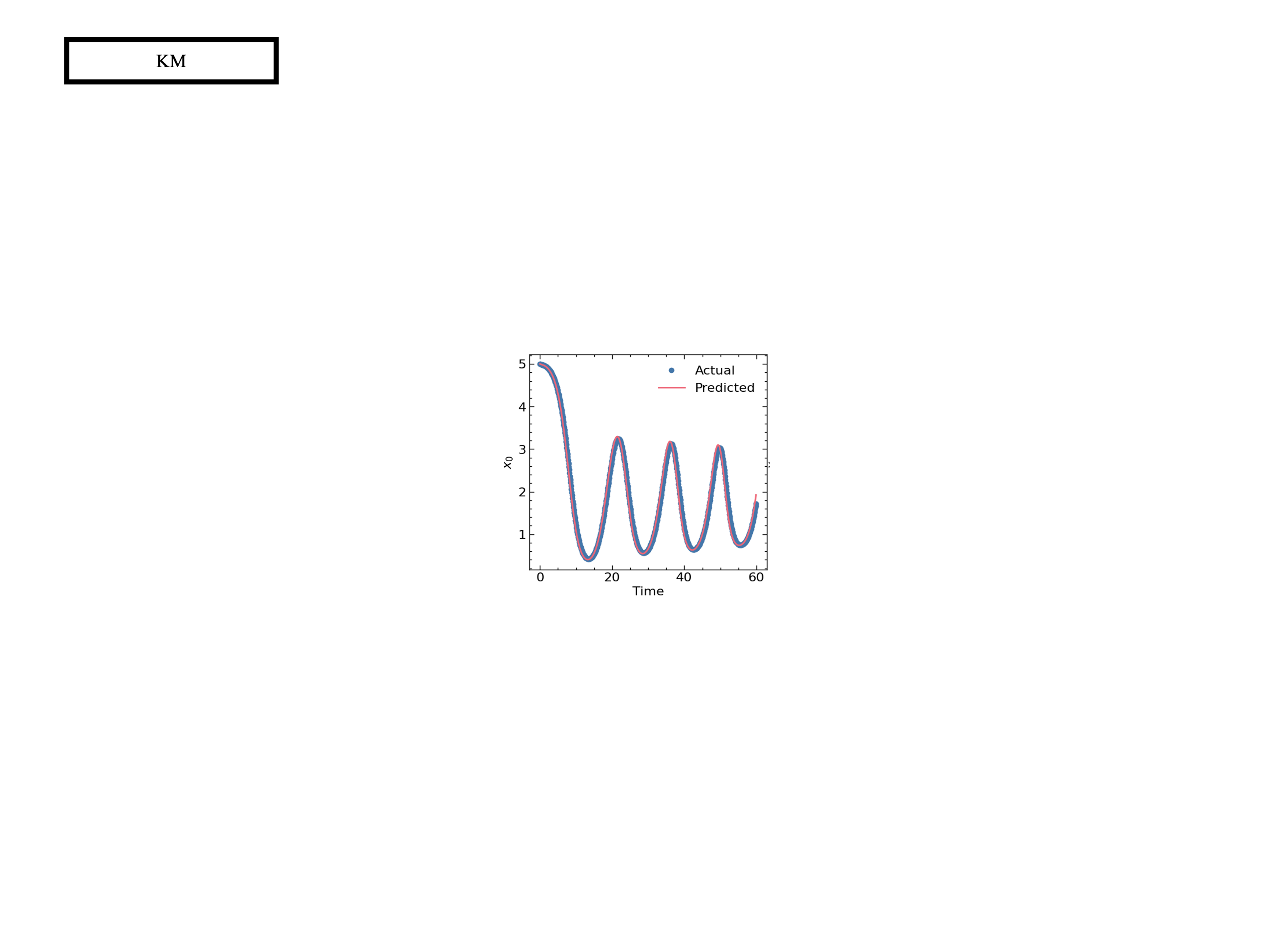}
      \caption{}
    \end{subfigure}
    \centering
    \begin{subfigure}{0.19\textwidth}
      \includegraphics[width = \linewidth, trim = 280 200 280 200, clip]{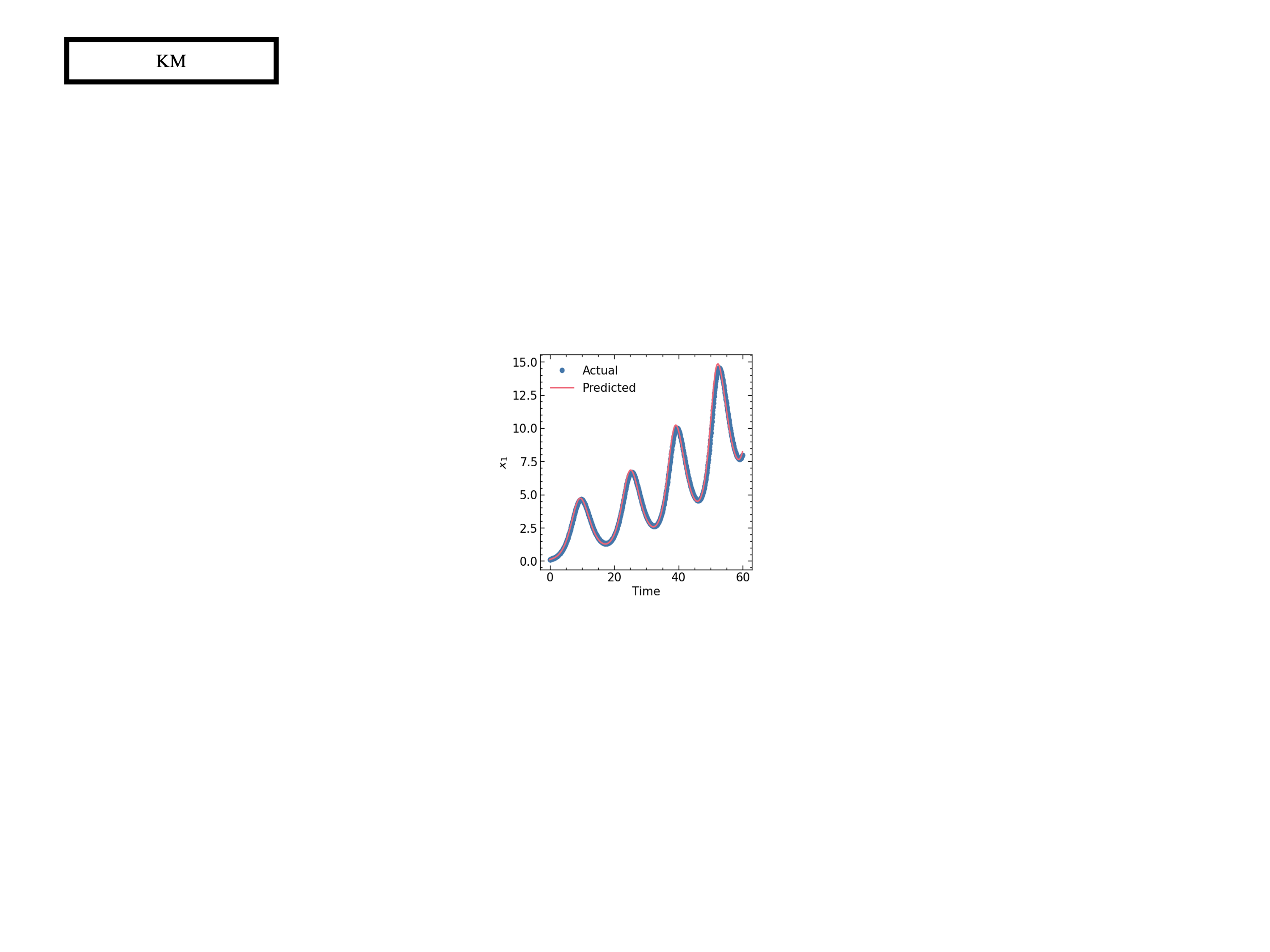}
      \caption{}
    \end{subfigure}
    \centering
    \begin{subfigure}{0.19\textwidth}
      \includegraphics[width = \linewidth, trim = 280 200 280 200, clip]{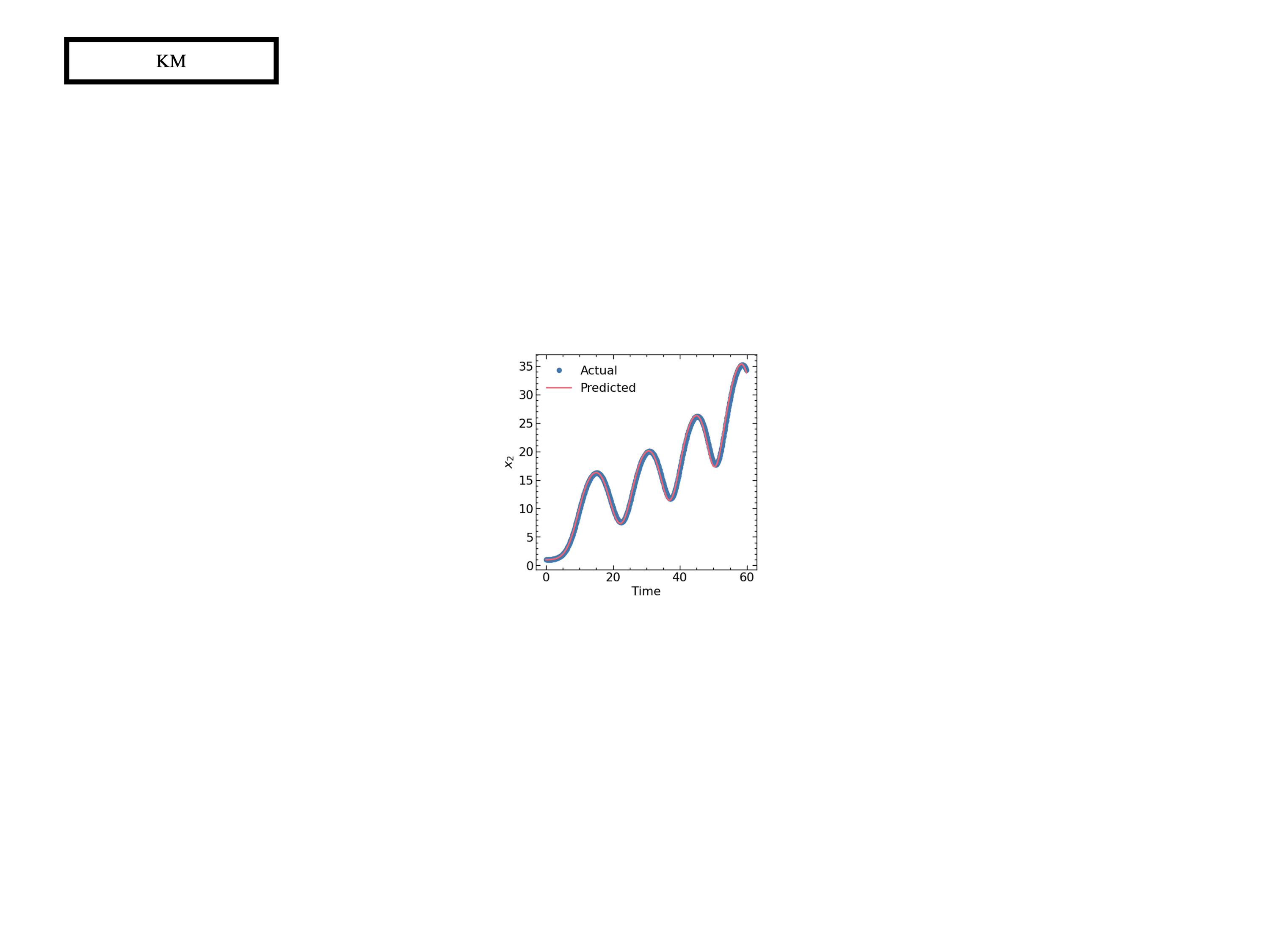}
      \caption{}
    \end{subfigure}
    \centering
    \begin{subfigure}{0.19\textwidth}
      \includegraphics[width = \linewidth, trim = 280 200 280 200, clip]{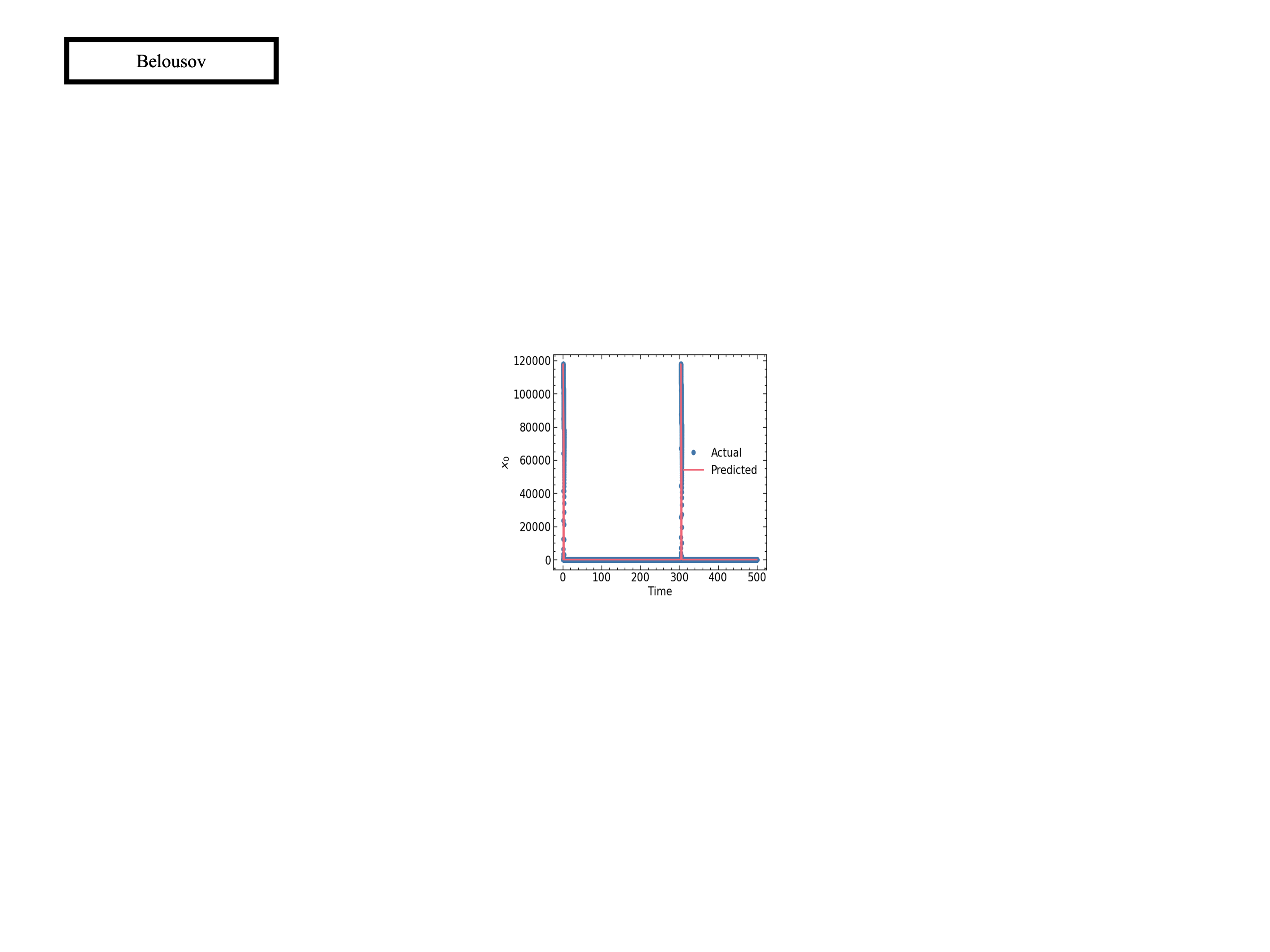}
      \caption{}
    \end{subfigure}
    \centering
    \begin{subfigure}{0.19\textwidth}
      \includegraphics[width = \linewidth, trim = 280 200 280 200, clip]{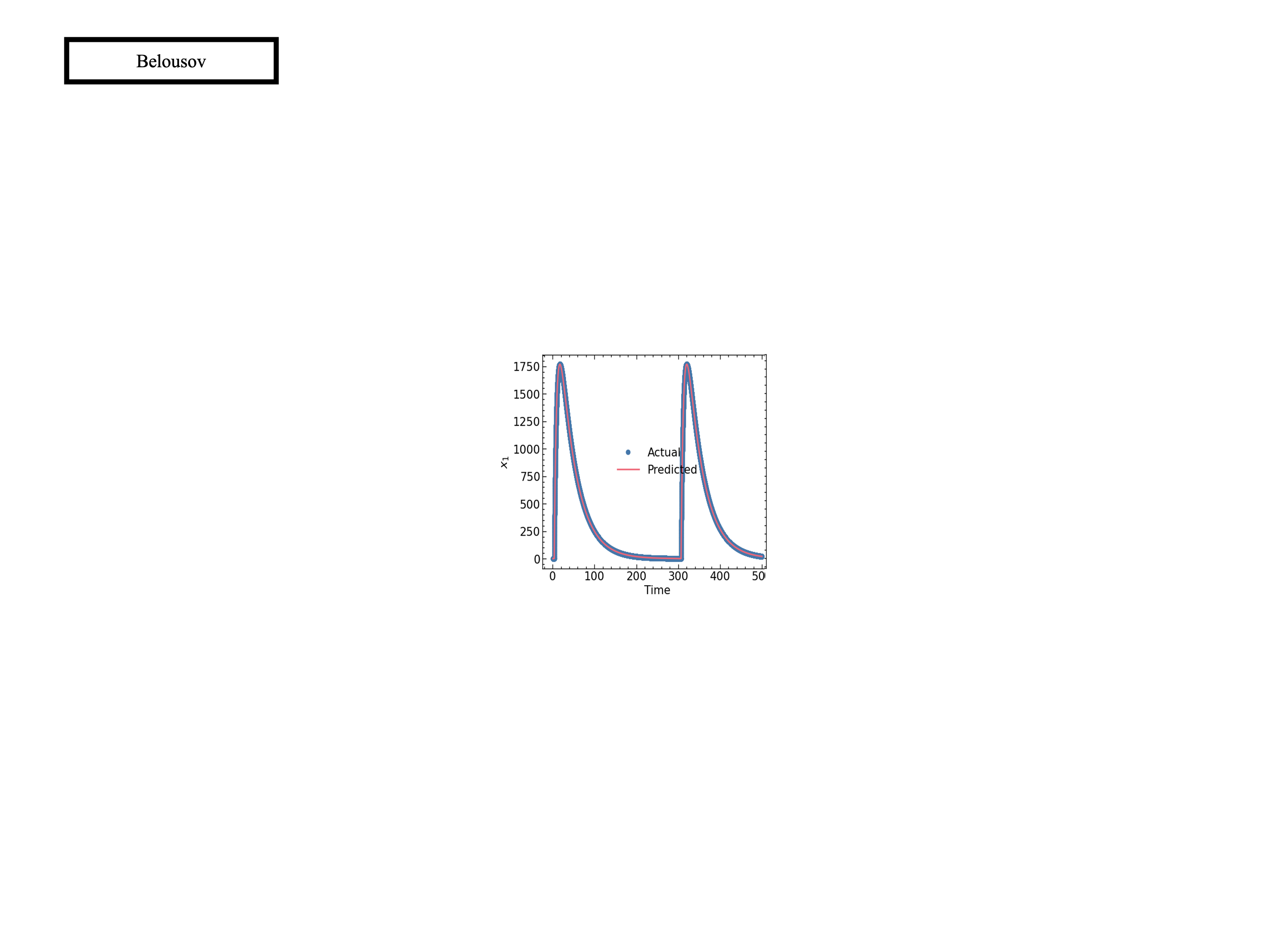}
      \caption{}
    \end{subfigure}
    \centering
    \begin{subfigure}{0.19\textwidth}
      \includegraphics[width = \linewidth, trim = 280 200 280 200, clip]{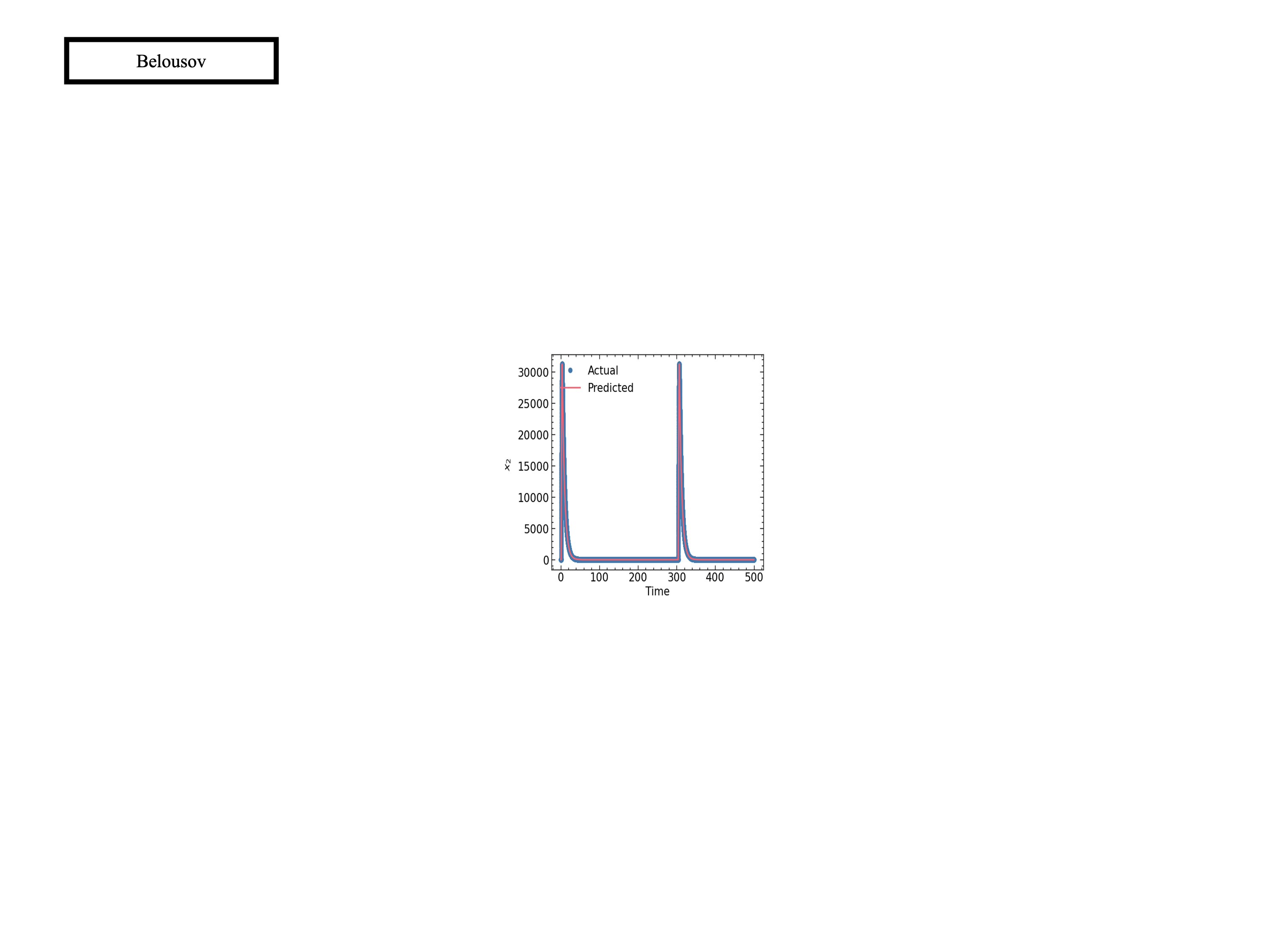}
      \caption{}
    \end{subfigure}
    \centering
    \begin{subfigure}{0.19\textwidth}
      \includegraphics[width = \linewidth, trim = 280 200 280 200, clip]{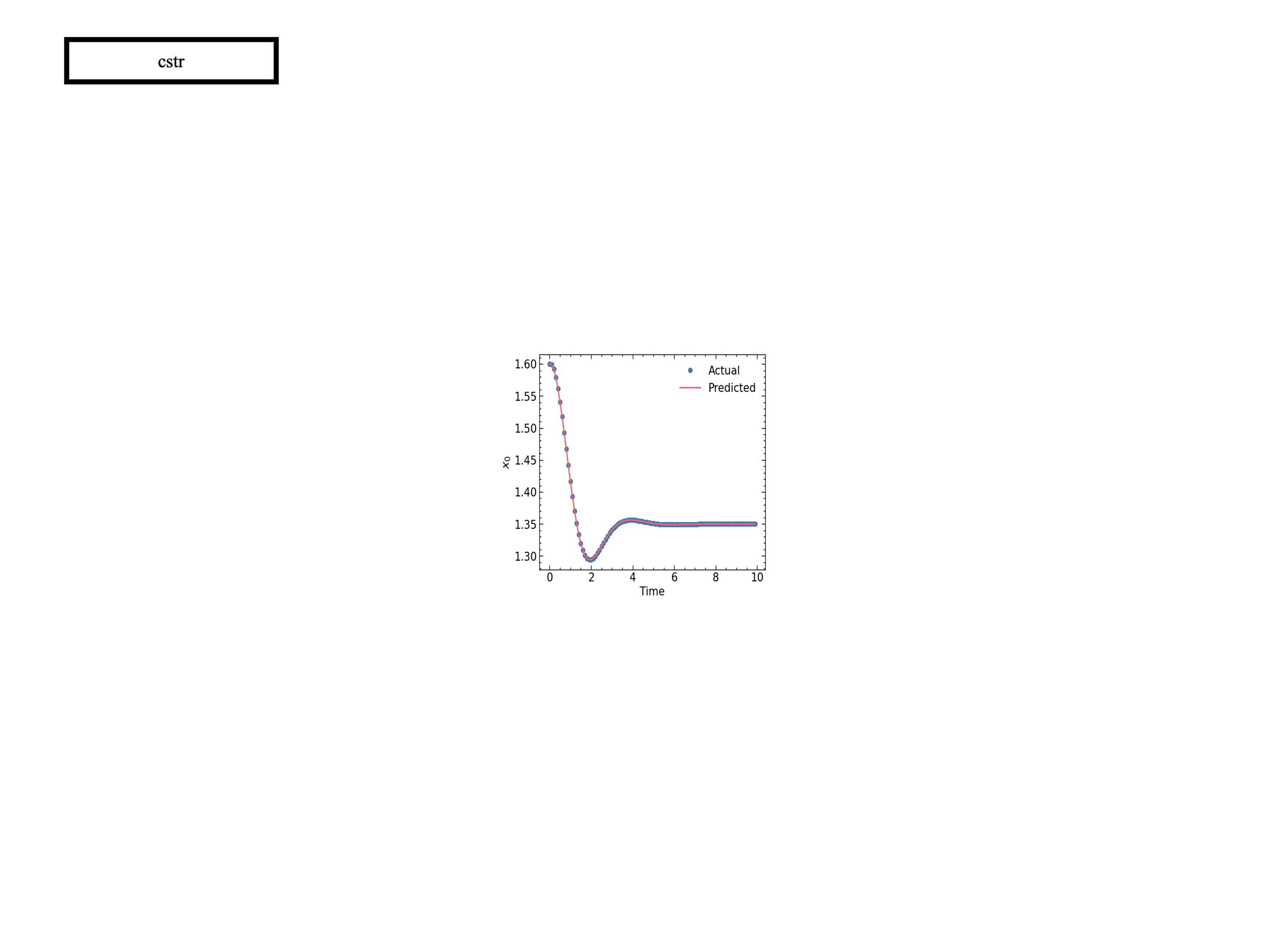}
      \caption{}
    \end{subfigure}
    \centering
    \begin{subfigure}{0.19\textwidth}
      \includegraphics[width = \linewidth, trim = 280 200 280 200, clip]{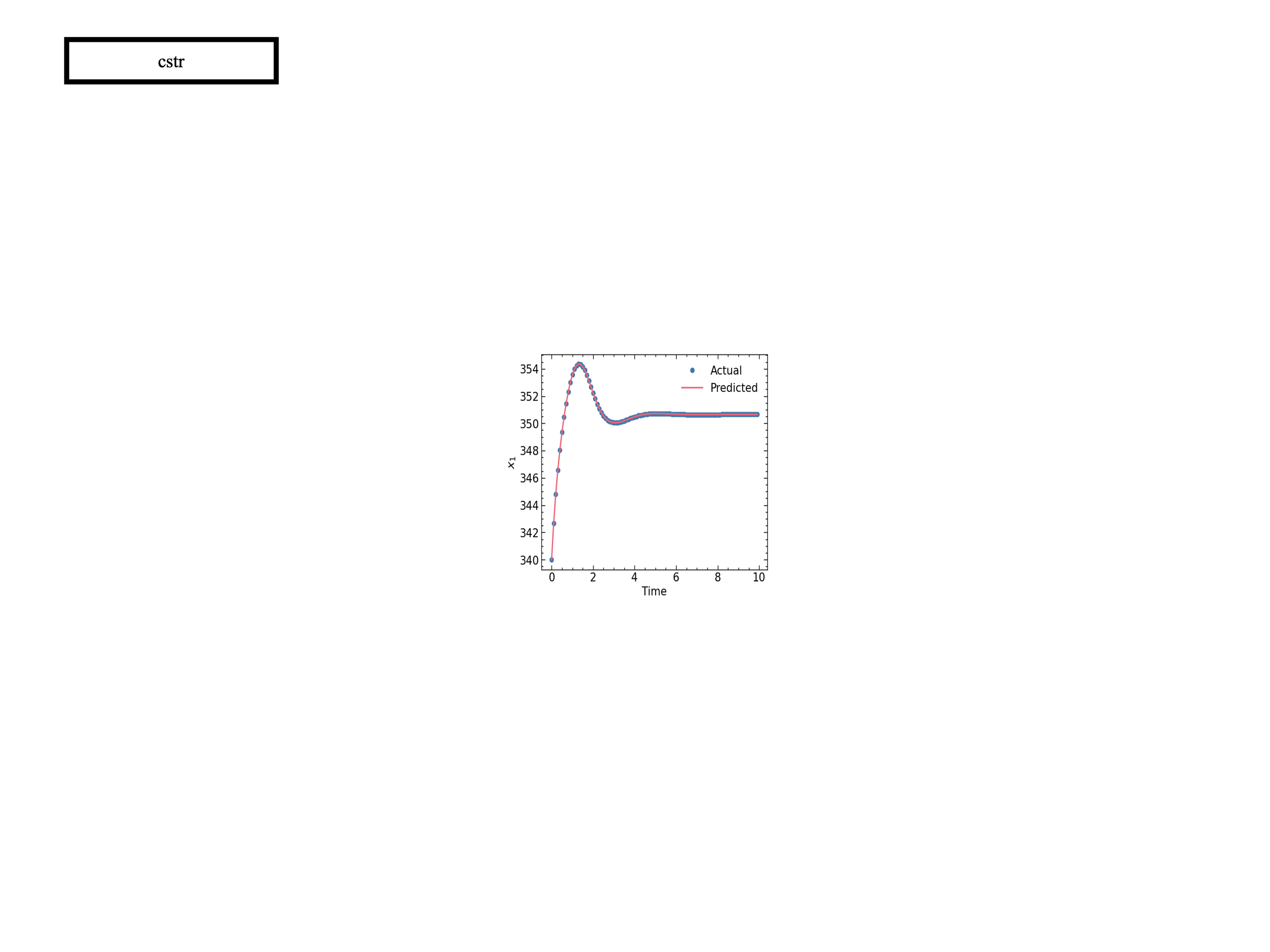}
      \caption{}
    \end{subfigure}
    \caption{A comparison of the experimental measurements (blue points) with the predicted state trajectories (solid lines) obtained by simulating Equation \ref{eqn:calcium} for the Calcium Ion system (a-d), Equation \ref{eqn:mendes} for the Mendes system (e-l), Equation \ref{eqn:ferment} for the Ethanol Fermentation system (m-p), Equation \ref{eqn:km} for the Kermack-McKendrick system (q-s), Equation \ref{eqn:br} for the Belousov reaction system (t-v), and Equation \ref{eqn:cstr} for the CSTR system (w-x), using the optimized parameters.}
    \label{fig:pstates}
\end{figure}

\begin{figure}[!ht]
    \centering
    \begin{subfigure}{\textwidth}
        \centering
        \includegraphics[width = \linewidth, height = 0.12\textheight]{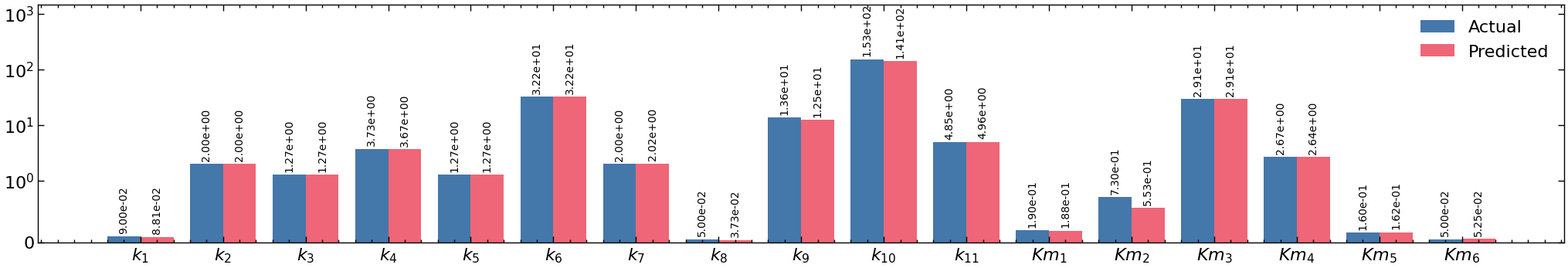}
        \caption{}
    \end{subfigure}
    \bigskip
    \centering
    \begin{subfigure}{\textwidth}
        \includegraphics[width = \linewidth, height = 0.2\textheight]{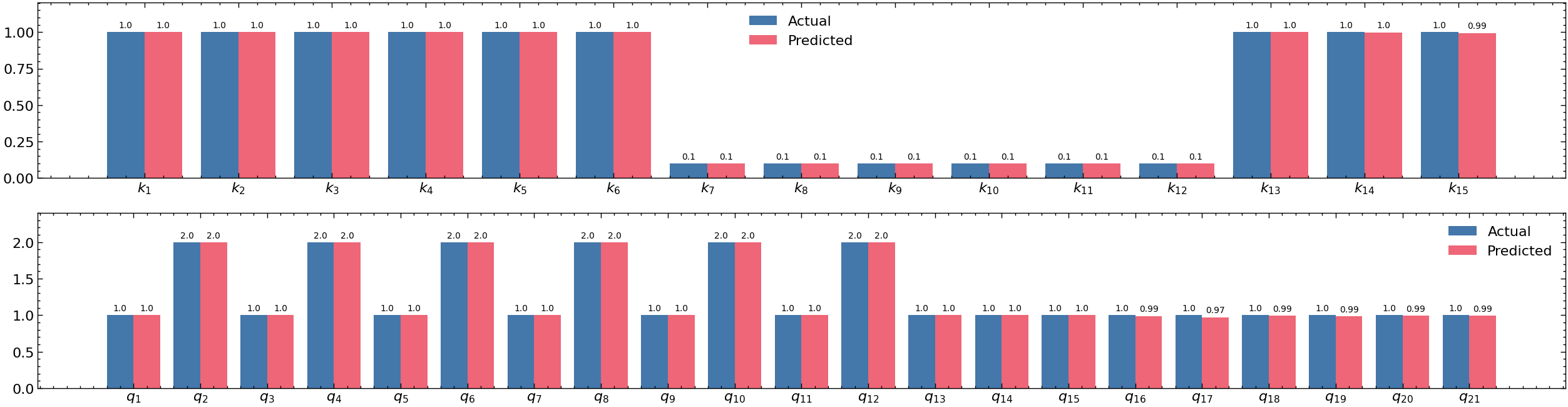}
        \caption{}
    \end{subfigure}
    \bigskip
    \centering
    \begin{subfigure}{\textwidth}
        \centering
        \includegraphics[width = \linewidth, height = 0.12\textheight]{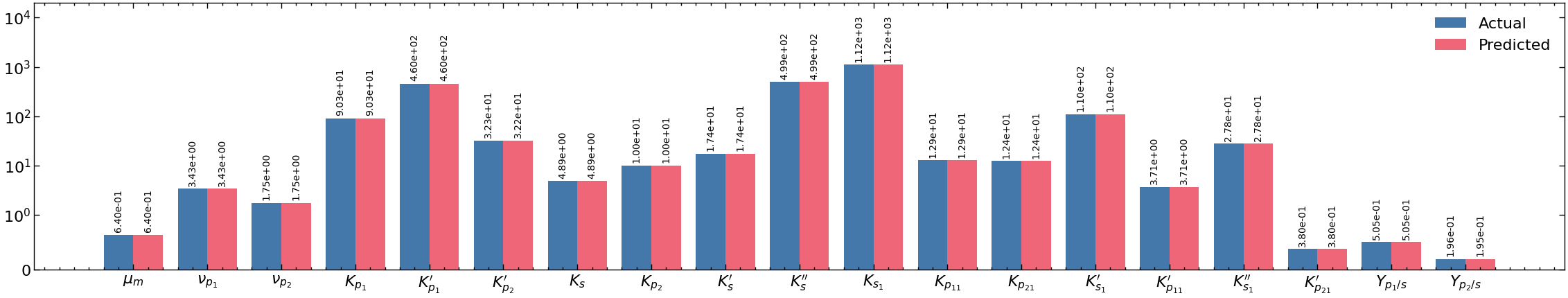}
        \caption{}
    \end{subfigure}
    \bigskip
    \centering
    \begin{subfigure}{0.5\textwidth}
        \centering
        \includegraphics[width = \linewidth, height = 0.12\textheight]{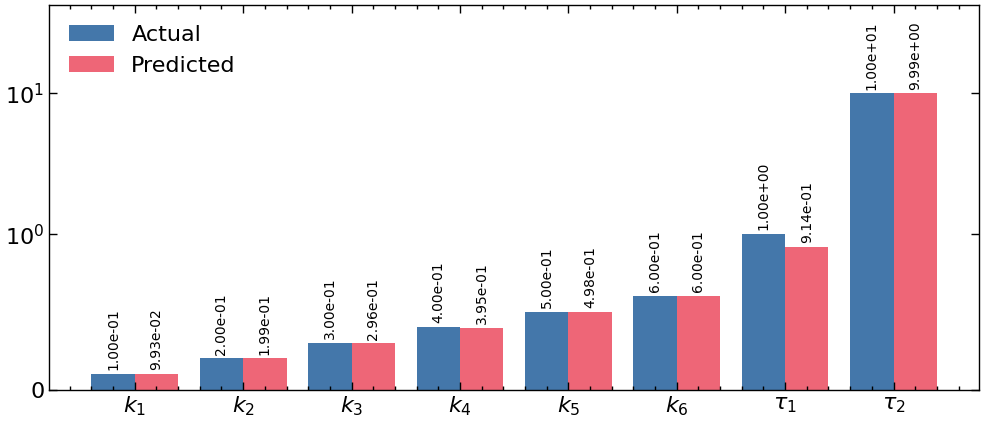}
        \caption{}
    \end{subfigure}
    \begin{subfigure}{0.3\textwidth}
        \centering
        \includegraphics[width = \linewidth, height = 0.12\textheight]{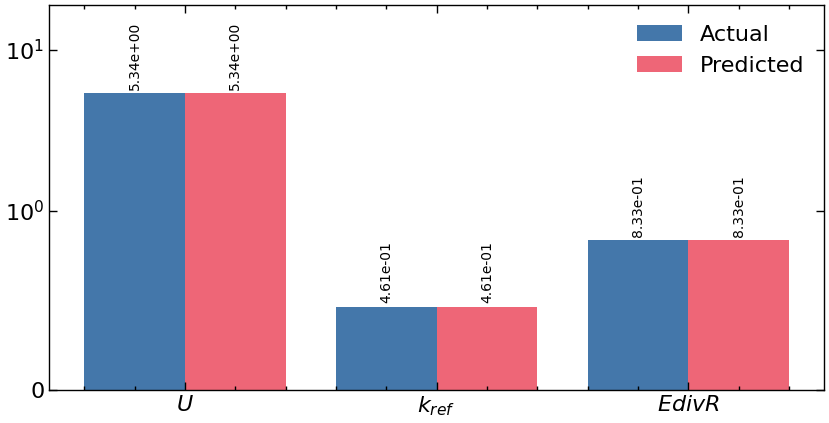}
        \caption{}
    \end{subfigure}
    \caption{A comparison of the actual (blue) and optimized (pink) linear and nonlinear parameters for Equation \ref{eqn:calcium} for the Calcium Ion system (a), Equation \ref{eqn:mendes} for the Mendes system (b), Equation \ref{eqn:ferment} for the Ethanol Fermentation system (c), Equation \ref{eqn:km} for the Kermack-McKendrick (d), and Equation \ref{eqn:cstr} for the CSTR system (e), respectively.}
    \label{fig:pparams}
\end{figure}

We observe that, for all benchmark problems mentioned in the table, the proposed bi-level optimization method converges to the globally optimal solution, whereas the single-shooting method and the orthogonal collocation method either fails to converge or converges only to a locally optimal solution under the same data and optimization budget. This behavior is also evident from Figure \ref{fig:pstates}, which compares the actual and predicted trajectories for all systems, and from Figure \ref{fig:pparams} and Table \ref{tab:param_carb}, which compare the true and estimated parameters for the parameter-estimation and model-discovery examples, respectively. This is likely because both single shooting and orthogonal collocation are highly sensitive to the initial parameter guess and, in the case of orthogonal collocation, to the initial states at the collocation points. Our proposed method, by contrast, exploits the convexity of the inner subproblem and is therefore able to achieve superior solutions consistently.

We further observe that the number of iterations required by our proposed approach is significantly smaller than that of the single-shooting method. This is expected, as our approach exploits convexity in the inner optimization problem. However, the wall time is substantially higher than that of the single-shooting approach. This is also expected because each outer iteration requires solving an inner convex optimization problem, and computing derivatives through the optimal solution is computationally expensive due to the need to invert the KKT matrix. Although some computational savings are possible by reusing the KKT matrix factorization provided by the optimization solver, not all solvers expose this information, and consequently the factorization must be recomputed during sensitivity calculations.

\begin{figure}[!ht]
    \centering
    \includegraphics[width = 0.4\linewidth, height = 0.22\textheight]{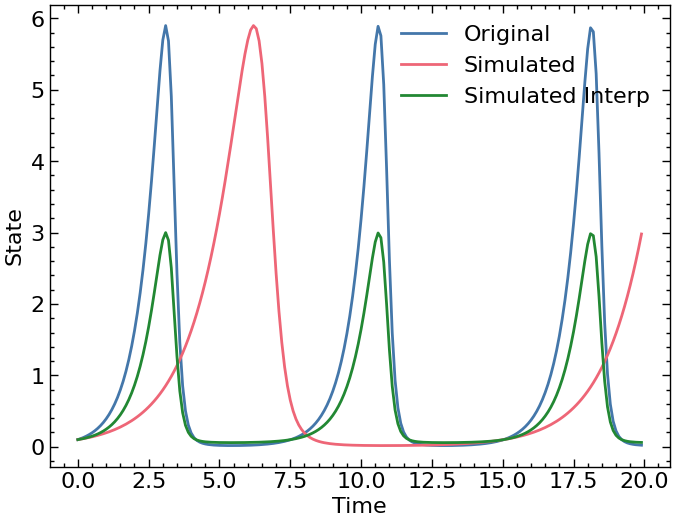}
    \caption{A comparison of the trajectories obtained without interpolation (red) and with interpolation (green) for a some parameters, alongside the original trajectory (blue) simulated using the optimal parameters.}
    \label{fig:stiff}
\end{figure}

For the Belousov reaction system in particular, single-shooting method fails to converge because the system is mildy stiff and an explicit solver is unable to adequately handle such stiffness. While this presents a challenge for single-shooting, an explicit solver can still be used within our proposed method. In single-shooting, the states are unknown during optimization, and a poor initial guess may steer the simulated trajectory toward undesired regions of the state space. In contrast, in our method the states are known and can be queried via interpolation on the measured data. This behavior is illustrated using a simple system in Figure \ref{fig:stiff}, where the original trajectory is shown in blue. For certain parameter values, simulating the system yields the trajectory shown in red, which represents the trajectory used by the single-shooting method. When interpolation is used instead, the trajectory in green remains closer to the original trajectory and is therefore less likely to diverge or exhibit numerical instability. Note that, although the sampling frequency used here is sufficient for identifying the correct parameters, parameter estimation for stiff systems using interpolation can be sensitive to the sampling frequency. In \cite{prabhu2025derivative}, we illustrate this effect using a small stiff example and analyze the impact of different sampling rates on parameter identification.

\begin{table}[ht]
\centering
\resizebox{\textwidth}{!}{%
\begin{tabular}{ccccccccccccc}
\hline
\multirow{2}{*}{System} & \multicolumn{4}{c}{Bi-Level} & \multicolumn{4}{c}{Single-Shooting} & \multicolumn{4}{c}{Orthogonal Collocation} \\\cmidrule{2-5} \cmidrule{6-9} \cmidrule{10-13}
& Status & $J$ & Iterations & \makecell[c]{Wall Time \\ (sec)} & Status & $J$ & Iterations & \makecell[c]{Wall Time \\ (sec)} & Status & $J$ & Iterations & \makecell[c]{Wall Time \\ (sec)}\\
\hline \\
Calcium Ion & Converged & $1.64e{-3}$ & 23 & 2183.91 & Failed & 2.16e{0} & 1000 & 1355.85 & Failed & 5.15e{1} & 1000 & 10669.82 \\
Mendes & Converged & $6.35e{-7}$ & 34 & 18124.8 & Failed & $1.63e{-1}$ & 113 & 267.83 & Failed & - & - & - \\
\makecell[c]{Ethanol \\ Fermentation} & Converged & $3.96e{-9}$ & 73 & 4873.67 & Failed & $1.4e{2}$ & 1000 & 417.67 & Failed & 2.16e{2} & 1000 & 768.62 \\
\makecell[c]{Belousov \\ Reaction} & Converged & $1.42e{2}$ & 15 & 14.76 & Failed & $3.67e{7}$ & 50 & 1061.66 & Failed & - & - & - \\
\hline \\
\end{tabular}%
}
\caption{Comparison of performance of the proposed bi-level optimization algorithm with single-shooting for parameter estimation of benchmark systems.}
\label{tab:results} 
\end{table}

\begin{table}[ht]
\centering
\begin{tabular}{ccccc}
\hline
\multirow{3}{*}{$R_i$} & \multicolumn{2}{c}{Forward reaction} & \multicolumn{2}{c}{Reverse reaction} \\
& Rate constant $(k)$ & Activation energy $( E \times 10^4)$ & Rate constant $(k)$ & Activation energy $(E \times 10^4)$\\
& $\left(\text{Actual}, \text{ Predicted} \right)$ & $\left(\text{Actual}, \text{ Predicted} \right)$ & $\left(\text{Actual}, \text{ Predicted} \right)$ & $\left(\text{Actual}, \text{ Predicted} \right)$ \\ 
\hline \\
1 & $  k^1_{43} = (0.3, \ 0.3)$ & $E^1_{43} = (9.48, 9.21) $ & $ k^1_{55} = (1.2, 1.2) $ & $ E^1_{55} = (8.58, 8.32) $ \\[4pt]
2 & $  k^2_{25} = (0.4, 0.4)$ & $E^2_{25} = (6.59 , 6.03 )$ & $k^2_{44} = (0.6, 0.59)$ & $ E^2_{44} = (2.77, 2.24)$\\[4pt]
3 & $ k^3_{17} = (1.1, 1.08)$ & $ E^3_{17} = (10.9, 9.49) $ & $k^3_{34} = (0.7, 0.68)$ & $E^3_{34} = (4.05, 2.67)$  \\[4pt]
4 & $ k^4_{24} = (0.9, 0.89)$ & $E^4_{24} = (4.88, 4.82)$ & $ k^1_{52} = (0.1, 0.1) $ & $ E^1_{52} = (5.59, 5.54) $  \\[4pt]
5 & $ k^5_{15} = (1, 0.97)$ & $ \ E^5_{15} = (7.74, 7.46)$ & $ k^1_{46} = (0.5, 0.48) $ & $ E^1_{46} = (1.14, 0.86) $  \\[4pt]
6 & $ k^6_{73} = (0.2, 0.2) $ & $  E^6_{73} = ( 10.9, 10.93 ) $ & $ k^6_{47} = (0.8, 0.79)$ & $ E^6_{47} = (3.46, 3.48)$ \\[4pt]
\hline \\
\end{tabular}
\caption{Summary of actual and predicted linear (reaction rate constants) and nonlinear parameters (activation energy) for each of the reactions in the esterification of carboxylic acid system.}
\label{tab:param_carb} 
\end{table}

\section{Conclusion}
In this work, we developed a bi-level optimization framework for parameter estimation of ordinary differential equation (ODE) systems. In this framework, an inner convex optimization problem is solved over the linear parameters, while the outer problem optimizes over the nonlinear parameters. We apply the implicit function theorem to the KKT conditions of the inner optimization problem to derive forward-mode sensitivities of the optimal solution with respect to the nonlinear parameters. Using several benchmark problems and two problem formulations—parameter estimation and model discovery—we demonstrate that the proposed framework not only successfully identifies the optimal parameters, but also converges in fewer iterations compared to sequential optimization (single-shooting).

While interpolation provides a promising alternative approach for parameter estimation in settings where data are abundant and frequently sampled, it also introduces certain limitations. Interpolation over noisy measurements, common in real-world, can lead to inaccurate integral approximations and, consequently, reduced solution accuracy. This issue can be mitigated to some extent through the use of smoothing techniques. Finally, interpolation based on sparsely sampled measurements may yield inaccurate integral estimates, in which case conventional sequential or simultaneous methods may be more suitable.

\section{Acknowledgements}
The authors acknowledge the financial support from Pennsylvania Infrastructure Technology Alliance. S.R. also acknowledges NSF CBET grant 2045550. Portions of this research were conducted on Lehigh University’s Research Computing infrastructure partially supported by NSF Award 2019035.

\clearpage
\appendix
\section{Parameter estimation benchmark problems}

In this section, we list the systems under consideration, including their governing equations, associated parameters, and data generation strategies.

\subsection{Calcium Ion Dynamics} \label{appendix:calcium}
The dynamics consist of four differential equations given as follows

\begin{equation}\label{eqn:calcium}
\begin{aligned}
    \frac{dx_0}{dt} & = k_1 + k_2 x_0 - k_3 x_1 \frac{x_0}{x_0 + Km_1} - k_4 x_2 \frac{x_0}{x_0 + Km_2} \\
    \frac{dx_1}{dt} & = k_5 x_0 - k_6 \frac{x_1}{x_1 + Km_3} \\
    \frac{dx_2}{dt} & = k_7 x_1 x_2 \frac{x_3}{x_3 + Km_4} + k_8 x_1 + k_9 x_0 - k_{10} \frac{x_2}{x_2 + Km_5} - k_{11} \frac{x_2}{x_2 + Km_6} \\
    \frac{dx_3}{dt} & = - k_7 x_1 x_2 \frac{x_3}{x_3 + Km_4} + k_{11} \frac{x_2}{x_2 + Km_6}
\end{aligned}
\end{equation}

\noindent where $x_0$, $x_1$, $x_2$, and $x_3$ are the concentration of four species, which interact in the
calcium-signaling pathway. The linear parameters ($p$) are $ k_1 = 0.09, k_2 = 2, k_3 = 1.27, k_4 = 3.73, k_5 = 1.27, k_6 = 32.24, k_7 = 2, k_8 = 0.05, k_9 = 13.58, k_{10} = 153, k_{11} = 4.85$ and the nonlinear parameters ($\phi$) are  $Km_1 = 0.19, Km_2 = 0.73, Km_3 = 29.09, Km_4 = 2.67, Km_5 = 0.16, Km_6 = 0.05$. The initial conditions are chosen to be $x_0(t = 0) = 0.12$, $x_1(t = 0) = 0.31$, $x_2(t = 0) = 0.0058$, $x_3(t = 0) = 4.3$. The model is simulated from $t_i = 0$ to $t_f = 20$(sec), and measurements are collected every $0.1$ seconds. For this set of parameters, the model exhibits a limit cycle. There are in total 17 parameters (11 appear linearly and 6 appear nonlinearly) to be estimated.

\subsection{Mendes Dynamics} \label{appendix:mendes}

This nonlinear biochemical dynamic model describes the variation of the metabolite concentrations with time. Consequently, $x_0$, $x_1$, $x_2$, $x_3$, $x_4$, $x_5$, $x_6$, and $x_7$ represent the concentrations of the species involved in the different biochemical reactions. There are 15 linear parameters ($p$), $k_{1-6} = 1, k_{7-12} = 0.1, k_{13-15} = 1$, and 21 nonlinear parameters ($\phi$), $q_{1,3,5,7,9,11,13-21} = 1, q_{2,4,6,8,10,12} = 2$. Thus, 36 total parameters need to be estimated. Furthermore, we assumed that all the linear parameters are greater than zero ($p \geq 0 $). We considered all $16$ combinations of $S \in \{ 0.1, 0.46416, 2.15, 10 \}$ and $P \in \{ 0.05, 0.13572, 0.3684, 1 \}$ and for each combination, we simulated the system from $t_i = 0$ to $t_f = 120$ (sec) and collected measurements every $0.1$ seconds. The initial conditions are $x_0 (t = 0) = 0.66667$, $x_1 (t = 0) = 0.57254$, $x_2  (t = 0) = 0.41758$, $x_3  (t = 0) = 0.4$, $x_4  (t = 0) = 0.36409$, $x_5  (t = 0) = 0.29457$, $x_6  (t = 0) = 1.419$, $x_7  (t = 0) = 0.93464$.

\begin{equation}\label{eqn:mendes}
\begin{aligned}
    \frac{dx_0}{dt} & = \frac{k_1}{1 + \left(\frac{P}{q_1}\right)^{q_2} + \left(\frac{q_3}{S}\right)^{q_4}} - k_2 x_0 \\
    \frac{dx_1}{dt} & = \frac{k_3}{1 + \left(\frac{P}{q_5}\right)^{q_6} + \left(\frac{q_7}{x_6}\right)^{q_8}} - k_4 x_1 \\
    \frac{dx_2}{dt} & = \frac{k_5}{1 + \left(\frac{P}{q_9}\right)^{q_{10}} + \left(\frac{q_{11}}{x_7}\right)^{q_{12}}} - k_6 x_2 \\
    \frac{dx_3}{dt} & = \frac{k_7 x_0}{x_0 + q_{13}} - k_8 x_3 \\
    \frac{dx_4}{dt} & = \frac{k_9 x_1}{x_1 + q_{14}} - k_{10} x_4 \\
    \frac{dx_5}{dt} & = \frac{k_{11} x_2}{x_2 + q_{15}} - k_{12} x_5 \\
    \frac{dx_6}{dt} & = \frac{k_{13} x_3 \left(\frac{1}{q_{16}}\right)(S - x_6)}{1 + \left(\frac{S}{q_{16}}\right) + \left(\frac{x_6}{q_{17}}\right)} - \frac{k_{14} x_4 \left(\frac{1}{q_{18}}\right)(x_6 - x_7)}{1 + \left(\frac{x_6}{q_{18}}\right) + \left(\frac{x_7}{q_{19}}\right)} \\
    \frac{dx_7}{dt} & = \frac{k_{14} x_4 \left(\frac{1}{q_{18}}\right)(x_6 - x_7)}{1 + \left(\frac{x_6}{q_{18}}\right) + \left(\frac{x_7}{q_{19}}\right)} - \frac{k_{15} x_5 \left(\frac{1}{q_{20}}\right)(x_7 - P)}{1 + \left(\frac{x_7}{q_{20}}\right) + \left(\frac{P}{q_{21}}\right)}
\end{aligned}
\end{equation}

\subsection{Ethanol Fermentation Dynamics}\label{appendix:ferment}

Equation \ref{eqn:ferment} describes the growth of microorganisms, the consumption of glucose and the formation of the products in a batch fermentation process. Here $x_0$ is the concentration of cell mass, $x_1$ is the concentration of glucose, $x_2$ is the concentration of ethanol, and $x_3$ is the concentration of glycerol. There are 3 linear parameters ($p$), $ \mu _m = 0.6397, \ \nu _{p_1} = 3.429, \ \nu _{p_2} = 1.748$ and 16 nonlinear parameters ($ \phi$), $ K_{p_1} = 90.35$, $K_{p_1}^{\prime} = 460.4$, $K_{p_2}^{\prime} = 32.26$, $K_s = 4.895$, $K_{p_2} = 10$, $K_s^{\prime} = 17.45$, $K_s^{\prime \prime} = 499.4$, $K_{s_1} = 1115.1$, $K_{p_{11}} = 12.89$, $K_{p_{21}} = 12.45$, $K_{s_1}^{\prime} = 110.3$, $K_{p_{11}}^{\prime} = 3.71$, $K_{s_1}^{\prime \prime} = 27.78$, $K_{p_{21}}^{\prime} = 0.3797$, $Y_{p_1/s} = 0.505$,  $Y_{p_2/s} = 0.1955$ that are estimated using measurements. Additionally, we assumed all the nonlinear parameters are greater than zero ($\phi \geq 0$). We simulated 15 experiments from $t_i = 0$ to $t_f = 15$(sec) and collected measurements every $0.1$ seconds. The initial conditions, for each of the experiments, were uniformly sampled for $x_0(t = 0) = [2, 10]$, $x_1 (t = 0) = [50, 100]$, $x_2 (t = 0) = [0, 10]$, $x_3 (t = 0) = [0, 10]$. 

\begin{equation}\label{eqn:ferment}
\begin{aligned}
    \frac{dx_0}{dt} & = x_0 \left( \frac{\mu _m x_1}{K_s + x_1 + x_1^2/K_{s_1}} \right) . \left( \frac{K_{p_1}}{K_{p_1} + x_2 + x_2^2/K_{p_{11}}} \right) .\left( \frac{K_{p_2}}{K_{p_2} + x_3 + x_3^2/K_{p_{21}}} \right) \\
    \frac{dx_1}{dt} & = \frac{x_0}{Y_{p_1/s}} \left( \frac{\nu _{p_1}x_1}{K_s^{\prime} + x_1 + x_1^2/K_{s_1}^{\prime}} \frac{K_{p_1}^{\prime}}{K_{p_1}^{\prime} + x_2 + x_2^2/K_{p_{11}}^{\prime}} \right) - \\ & \quad \quad \quad \frac{x_0}{Y_{p_2/s}} \left( \frac{\nu _{p_2}x_1}{K_s^{\prime \prime} + x_1 + x_1^2/K_{s_1}^{\prime \prime}} \frac{K_{p_2}^{\prime}}{K_{p_2}^{\prime} + x_3 + x_3^2/K_{p_{21}}^{\prime}} \right) \\
    \frac{dx_2}{dt} & = x_0 \left( \frac{\nu _{p_1}x_1}{K_s^{\prime} + x_1 + x_1^2/K_{s_1}^{\prime}} \right) . \left( \frac{K_{p_1}^{\prime}}{K_{p_1}^{\prime} + x_2 + x_2^2/K_{p_{11}}^{\prime}} \right)\\
    \frac{dx_3}{dt} & = x_0 \left( \frac{\nu _{p_2}x_1}{K_s^{\prime \prime} + x_1 + x_1^2/K_{s_1}^{\prime \prime}} \right) . \left( \frac{K_{p_2}^{\prime}}{K_{p_2}^{\prime} + x_3 + x_3^2/K_{p_{21}}^{\prime}} \right)\\
\end{aligned}
\end{equation}

\subsection{Kermack-McKendrick Dynamics}\label{appendix:km}

The Kermack-McKendrick system is given by the following delayed differential equation

\begin{equation}\label{eqn:km}
\begin{aligned}
    \frac{dx_0}{dt} & = - k_1 x_0 x_1(t - \tau _1) + k_2 x_1 (t - \tau_2) \\
    \frac{dx_1}{dt} & = k_3 x_0 x_1(t - \tau _1) - k_4 x_1 \\
    \frac{dx_2}{dt} & = k_5 x_1 - k_6 x_1(t - \tau_2)
\end{aligned}
\end{equation}

\noindent where $x_0$, $x_1$, and $x_2$ are the number of susceptible people, the number of infected people, and the number of people who have recovered and developed immunity to the infection, respectively. $x_1(t - \tau_1)$ represents the value of the state $x_1$ at time $(t - \tau_1)$. The linear parameters ($p$) are $k_1 = 0.1, k_2 = 0.2, k_3 = 0.3, k_4 = 0.4, k_5 = 0.5, k_6 = 0.6$ and the nonlinear parameters ($\phi$) are $\tau _1 = 1, \tau _2 = 10$. We simulated the system from $t_i = 0$ to $t_f = 60$(sec) and collected measurements every $0.1$ seconds. The initial conditions are specified as $x_0(t \leq 0) = 5$, $x_1(t \leq 0) = 0.1$, $x_2(t \leq 0) = 1$. 

\subsection{Belousov Reaction Dynamics}\label{appendix:br}

We consider the following dynamic equations for the Belousov reaction in a continuously stirred batch reactor.

\begin{equation}\label{eqn:br}
\begin{aligned}
    \frac{dx_0}{dt} & = k_1 (x_1 + k_4 x_0 - x_1 x_0 - k_5 x_0^2) \\
    \frac{dx_1}{dt} & = \frac{- k_2 x_1 - x_0 x_1 + x_2}{k_1} \\
    \frac{dx_2}{dt} & = k_3 (x_0 - x_2)
\end{aligned}
\end{equation}

\noindent where $x_0$, $x_1$, $x_2$ are the concentration of species in the reactor. The linear parameters ($p$) are $k_2 = 1$, $k_3 = 0.161$, $k_4 = 1$, $k_5 = 8.375\times 10^{-6}$ and the nonlinear ($\phi$) parameter is $k_1 = 77.27$. The system is integrated from $t_i = 0$ to $t_f = 500$(sec) with measurement interval $0.01$ seconds and with initial conditions uniformly sampled for $x_i(t = 0) = [0.1, 0.5], i \in \{0, 1, 2\}$.   

\subsection{Continuously Stirred Tank Reactor Dynamics}\label{appendix:cstr}

We demonstrate using a simple example of a continuously stirred tank reactor (CSTR), in which an exothermic reaction takes place. The dynamics of the system are given as   

\begin{equation}\label{eqn:cstr}
\begin{aligned}
    \frac{dx_0}{dt} & = - \hat{k} x_0 + F_{\text{in}} (C_{\text{in}} - x_0) \\
    \frac{dx_1}{dt} & = 130 \hat{k}x_0 + F_{\text{in}} (T_{\text{in}} - x_1) + U (T_c - x_1) \\
    \hat{k} & = k_{ref} e^{-10^4 EdivR \left( \frac{1}{x_1} - \frac{1}{350}\right)}
\end{aligned}
\end{equation}

\noindent where $x_0$ and $x_1$ are the concentration (unobserved state) and the temperature (observed state) inside the reactor. $C_{\text{in}}, \ F_{\text{in}}, \ \text{and} \ T_{\text{in}}$ are the concentration, flowrate, and temperature of the inlet stream. The nonlinear parameters ($\phi$) are given as $k_{ref} = 0.461$, $EdivR = 0.833$, while the linear parameter ($p$) is given as $U = 5.3417$. Note that although $EdivR$ is nonlinear, $k_{ref}$ could have been treated as linear, however, since $x_0$ is unmeasured, all parameters that appear in the coefficient of $x_0$ become nonlinear. The model is simulated from $t_i = 0$ to $t_f = 15$(sec), and measurements are collected every $0.1$ seconds. Although the concentration is unobserved, the initial concentration is known to be $x_0 = 1.6$, and the initial temperature is $x_1 = 340$. 

\subsection{Esterification of Carboxylic Acid Dynamics}\label{appendix:carb}

The overall reaction network of esterificaiton of carboxylic acid \cite{https://doi.org/10.1002/aic.690471016} is summarized in Table \ref{tab:reaction_carboxylic}. This reaction network consists of six reversible reactions and 11 different species, the reaction rates of which are given in Equation \ref{eqn:carb}

\begin{table}[ht]
\centering
\begin{tabular}{cccc}
\hline
\multicolumn{1}{c}{$R_i$} & \multicolumn{1}{c}{Reaction} & \multicolumn{1}{c}{Forward reaction} & \multicolumn{1}{c}{Reverse reaction} \\
\hline \\
1 &  HX $+$ H$_2$O $\longleftrightarrow$ H$_3$O$^+$ $+$ X & $ k_1 = 0.3, \ E_1 = 9.48 \times 10^4$ & $ k_7 = 1.2, \ E_7 = 8.58 \times 10^4 $ \\
2 &  H$_3$O$^+$ $+$ RCOOH $\longleftrightarrow$ RC$^+$(OH)$_2$ $+$ H$_2$O & $ k_2 = 0.4, \ E_2 = 6.59 \times 10^4 $ & $ k_8 = 0.6, \ E_8 = 2.77 \times 10^4 $ \\
3 &  R$^{'}$OH $+$ G $\longleftrightarrow$ H$_3$O$^+$ $+$ RCOOR$^{'}$ & $ k_3 = 1.1, \ E_3 = 10.9 \times 10^4 $ & $ k_9 = 0.7, \ E_9 = 4.05 \times 10^4 $ \\
4 &  RCOOH $+$ H$_2$O $\longleftrightarrow$ RCOO$^-$ $+$ H$_3$O$^+$ & $ k_4 = 0.9, \ E_4 = 4.88 \times 10^4 $ & $ k_{10} = 0.1, \ E_{10} = 5.59 \times 10^4$ \\
5 &  R$^{'}$OH $+$ H$_3$O$^+$ $\longleftrightarrow$ H$_2$O $+$ R$^{'}$OH$_2^+$ & $ k_5 = 1.0, \ E_5 = 7.74 \times 10^4 $ & $ k_{11} = 0.5, \ E_{11} = 1.14 \times 10^4 $ \\
6 &  X $+$ R$^{'}$OH$_2^+$ $\longleftrightarrow$ R$^{'}$X $+$ H$_2$O & $ k_6 = 0.2, \ E_6 = 10.9 \times 10^4 $ & $ k_{12} = 0.8, \ E_{12} = 3.46 \times 10^4 $ \\ 
\hline \\
\end{tabular}
\caption{Reaction mechanism of esterification of carboxylic acid}
\label{tab:reaction_carboxylic} 
\end{table}

\begin{equation}\label{eqn:carb}
\begin{aligned}
    \frac{d}{dt}\begin{bmatrix}
        x_0 \\
        x_1 \\
        x_2 \\
        x_3 \\
        x_4 \\
        x_5 \\
        x_6 \\
        x_7 \\
        x_8 \\
        x_9 \\
        x_{10} \\
    \end{bmatrix} = 
    \begin{bmatrix}
     0 &  0 & -1 &  0 & -1 &  0 \\
     0 & -1 &  0 & -1 &  0 &  0 \\
     0 &  0 &  1 &  0 &  0 &  0 \\
    -1 &  1 &  0 & -1 &  1 &  1 \\
     1 & -1 &  1 &  1 & -1 &  0 \\
    -1 &  0 &  0 &  0 &  0 &  0 \\
     0 &  1 & -1 &  0 &  0 &  0 \\
     0 &  0 &  0 &  1 &  0 &  0 \\
     0 &  0 &  0 &  0 &  1 & -1 \\
     0 &  0 &  0 &  0 &  0 &  1 \\
     1 &  0 &  0 &  0 &  0 & -1 \\
    \end{bmatrix}_{11 \times 6}
    \begin{bmatrix}
    \hat{k}_1 x_3 x_5 - \hat{k}_7 x_4 x_{10} \\[4pt]
    \hat{k}_2 x_1 x_4 - \hat{k}_8 x_3 x_6 \\[4pt]
    \hat{k}_3 x_0 x_6 - \hat{k}_9 x_2 x_4 \\[4pt]
    \hat{k}_4 x_1 x_3 - \hat{k}_{10} x_4 x_7 \\[4pt]
    \hat{k}_5 x_0 x_4 - \hat{k}_{11} x_3 x_8 \\[4pt]
    \hat{k}_6 x_8 x_{10} - \hat{k}_{12} x_3 x_9 \\[4pt]
    \end{bmatrix} \begin{matrix}
        \rightarrow R_1 \\[4pt]
        \rightarrow R_2 \\[4pt]
        \rightarrow R_3 \\[4pt]
        \rightarrow R_4 \\[4pt]
        \rightarrow R_5 \\[4pt]
        \rightarrow R_6 \\[4pt]
    \end{matrix}
\end{aligned}
\end{equation}

where $x_0, x_1, x_2, x_3, x_4, x_5, x_6, x_7, x_8, x_9, x_{10}$ are the concentrations of species R$^{'}$OH, RCOOH, RCOOR$^{'}$, H$_2$O, H$_3$O$^+$, HX, RC$^+$(OH)$_2$, RCOO$^-$, R$^{'}$OH$_2^+$, R$^{'}$X, X respectively. The reaction rate constant $\hat{k}_i$ is temperature ($T$) dependent and is given by the Arrhenius equation. The reaction rates at reference temperature of $373$K, denoted by $k_i$ (linear parameters $p$) and the activation energies, denoted by $E_i$ ( nonlinear parameters) are summarized for each of the reactions in Table \ref{tab:reaction_carboxylic}. The equations are simulated from $t_i = 0$ to $t_f = 10$ (sec) and measurements are collected every $0.05$ seconds. We consider 30 independent experiments for which the initial conditions are randomly chosen ($x(t = 0) \sim \mathcal{U}(4, 10)$) and the temperature chosen as $\{ 370, 375, 380, 385, 373 \}$. In this problem, we want to estimate $k_i$, $E_i$ and the functional form of $\Theta$. We assume that $\Theta$ is an additive combination of a subset of basis functions drawn from a library of all possible polynomial combinations (upto degree 2) of states \citep{brunton2016discovering} given by Equation \ref{eqn:library}. Additionally, each additive term in Equation \ref{eqn:library} has its temperature-dependent reaction rate constant given by the Arrhenius equation. Thus, the problem boils down to selecting the correct terms for each of the reactions ($R_i$) and estimating their coefficients.

\begin{equation}\label{eqn:library}
\begin{aligned}
    \frac{d}{dt}\begin{bmatrix}
        x_0 \\
        x_1 \\
        x_2 \\
        x_3 \\
        x_4 \\
        x_5 \\
        x_6 \\
        x_7 \\
        x_8 \\
        x_9 \\
        x_{10} \\
    \end{bmatrix} & = 
    \underbrace{\begin{bmatrix}
     0 &  0 & -1 &  0 & -1 &  0 \\
     0 & -1 &  0 & -1 &  0 &  0 \\
     0 &  0 &  1 &  0 &  0 &  0 \\
    -1 &  1 &  0 & -1 &  1 &  1 \\
     1 & -1 &  1 &  1 & -1 &  0 \\
    -1 &  0 &  0 &  0 &  0 &  0 \\
     0 &  1 & -1 &  0 &  0 &  0 \\
     0 &  0 &  0 &  1 &  0 &  0 \\
     0 &  0 &  0 &  0 &  1 & -1 \\
     0 &  0 &  0 &  0 &  0 &  1 \\
     1 &  0 &  0 &  0 &  0 & -1 \\
    \end{bmatrix}}_{\text{Stoichiometric Matrix}} \underbrace{
    \begin{bmatrix}
    \sum _{i = 0}^{10} \hat{k}_i^1x_{i} + \sum _{i = 0}^{10} \sum_{j = i}^{10} \hat{k}_{(i, j)}^1 x_{i}x_j \\[6pt]
    \sum _{i = 0}^{10} \hat{k}_i^2x_{i} + \sum _{i = 0}^{10} \sum_{j = i}^{10} \hat{k}_{(i, j)}^2 x_{i}x_j \\[6pt]
    \sum _{i = 0}^{10} \hat{k}_i^3x_{i} + \sum _{i = 0}^{10} \sum_{j = i}^{10} \hat{k}_{(i, j)}^3 x_{i}x_j \\[6pt]
    \sum _{i = 0}^{10} \hat{k}_i^4x_{i} + \sum _{i = 0}^{10} \sum_{j = i}^{10} \hat{k}_{(i, j)}^4 x_{i}x_j \\[6pt]
    \sum _{i = 0}^{10} \hat{k}_i^5x_{i} + \sum _{i = 0}^{10} \sum_{j = i}^{10} \hat{k}_{(i, j)}^5 x_{i}x_j \\[6pt]
    \sum _{i = 0}^{10} \hat{k}_i^6x_{i} + \sum _{i = 0}^{10} \sum_{j = i}^{10} \hat{k}_{(i, j)}^6 x_{i}x_j \\[6pt]
    \end{bmatrix}}_{\text{Polynomial Terms}} \begin{matrix}
        \rightarrow R_1 \\[6pt]
        \rightarrow R_2 \\[6pt]
        \rightarrow R_3 \\[6pt]
        \rightarrow R_4 \\[6pt]
        \rightarrow R_5 \\[6pt]
        \rightarrow R_6 \\[6pt]
    \end{matrix}\\
    \quad \text{Indexing }(i, j) & = 11 + \frac{(23 - i)i}{2} + (j - i), \quad \text{Arrhenius Equation } \hat{k}^n_{(i, j)} = k^n_{(i, j)} e^{\frac{-E^n_{(i, j)}}{R}\left( \frac{1}{T} - \frac{1}{373} \right)} \ 
\end{aligned}
\end{equation}

\clearpage
\section{Parameter estimation as an optimization problem}\label{appendix:opti}
In this section, we formulate the optimization problems underlying two standard parameter estimation methods for ordinary differential equations, namely single shooting and orthogonal collocation, which serve as baselines for comparison with the proposed approach. A simple Python implementation of the Lotka-Volterra system is also provided.

\subsection{Single Shooting}

The single-shooting method is a classical approach for parameter estimation in differential equations, where an optimization problem is formulated with unknown parameters as decision variables. At each iteration of the optimization solve, the ODE system is integrated forward in time from a given initial condition using a numerical solver, and the resulting trajectory is compared against the observed data through a least-squares or maximum likelihood objective. Any constraints on the parameters are handled by the optimization solver. The gradients (sensitivities) with respect to the parameters, required by the optimization solver, must be computed across the numerical integrator. While conceptually straightforward and easy to implement, the method suffers from several well-known drawbacks, including high computational cost due to repeated numerical integration, sensitivity to the initial parameter guess, and numerical instability for stiff or chaotic systems where small perturbations in parameters can lead to large deviations in the integrated trajectory. This is summarized as solving the following optimization problem

\begin{align}
\begin{split}
    \min _{p} L(p) = & \ \frac{1}{2} \sum_{t = t_i}^{t_f} || \hat{X}(t) - \hat{X}(0) - \int _0^t f(X, p) ||^2  \\
    \text{subject to} & \\
    & g(p) = 0 \\
    & h(p) \leq 0
\end{split}
\end{align}

\noindent where $p$ are the unknown parameters, $L$ is the mean squared error loss function, $\hat{X}$ are the measurements of states and $f$ is the system of differential equations. $g$ and $h$ are the corresponding equality and inequality constraints. A simple code example of estimating the parameters of Lotka-Volterra system is shown below

\begin{lstlisting}[style=pythonstyle]
    import jax
    import jax.numpy as jnp
    jax.config.update("jax_enable_x64", True)
    import diffrax
    from cyipopt import minimize_ipopt
    
    
    # Forward and reverse mode autodiff compatible ODE solver
    def odeint_diffrax(
            afunc, xinit, time_span, parameters, 
            rtol = 1e-6, atol = 1e-8, mxstep = 10_000
        ) :
        # Forward and reverse mode autodiff compatible ode solver
        _afunc = lambda t, x, p : afunc(x, t, p)
        return diffrax.diffeqsolve(
                    diffrax.ODETerm(_afunc), 
                    diffrax.Tsit5(),
                    t0 = time_span[0], 
                    t1 = time_span[-1],
                    dt0 = None, 
                    saveat = diffrax.SaveAt(ts = time_span), 
                    y0 = xinit, 
                    args = parameters,
                    stepsize_controller = diffrax.PIDController(
                        rtol=rtol, atol=atol, 
                        pcoeff = 0.4, icoeff = 0.3, dcoeff = 0.
                    ),
                    adjoint = diffrax.DirectAdjoint(), 
                    max_steps = mxstep
            ).ys
    
    # Governing equations
    def LotkaVolterra(x, t, p):
        a, b, c, d = p
        return jnp.array([
            a * x[0] - b * x[0] * x[1],
            - c * x[1] + d * x[0] * x[1]
        ])
    
    # Generate data
    t0, tf, dt = 1, 10, 0.1 
    time_span = jnp.arange(t0, tf, dt)
    xinit = jnp.array([0.1, 0.1])
    p_actual = jnp.array([2/3, 4/3, 1, 1])
    solution = odeint_diffrax(LotkaVolterra, xinit, time_span, p_actual)

    # Define objective function
    def objective_function(p):
        prediction = odeint_diffrax(LotkaVolterra, xinit, time_span, p)    
        return jnp.mean((prediction - solution)**2)
    
    # Set solver options and solve the problem
    results = minimize_ipopt(
        jax.jit(objective_function), # JIT compiled objective 
        x0 = jnp.zeros(4), # intial guess 
        jac = jax.jit(jax.grad(objective_function)), # JIT compiled Jacobian
        hess = jax.jit(jax.hessian(objective_function)), # JIT compiled Hessian
        tol = 1e-8, 
        options = {"maxiter" : 1000, "disp" : 5,}
        )
    
    p_opt = jnp.array(results.x) # optimal parameters
\end{lstlisting}

\subsection{Orthogonal Collocation}

Collocation over finite elements, often referred to as collocation, is another widely used numerical method for discretizing differential equations. The method begins by partitioning the continuous domain into $N-1$ finite elements. Within each finite element $ i $, the state $x(t)$ is approximated by a polynomial of order $K + 1$, defined at $K$ collocation points. 

\begin{equation}
\begin{aligned}
    x(t) & = \sum_{j = 0}^{K} l_j(\tau)x_{ij}, \quad t \in [t_{i - 1}, t_i], \quad \tau \in [0, 1] \\
    \text{where} \\
    l_j(\tau) & = \prod _{k = 0, \neq j}^K \frac{\tau - \tau_k}{\tau_j - \tau_k}
\end{aligned}
\end{equation}

In addition to the unknown parameters, the values of the state at the collocation points also enter as decision variables. Given this polynomial approximation, the differential equation can be enforced as a constraint at each collocation point within each finite element. Continuity of the state between finite elements is enforced as additional equality constraints at the boundaries.

\begin{equation}
\begin{aligned}
    \frac{1}{h_i} \sum_{j = 0}^K x_{ij} \frac{dl_j(\tau_k)}{d\tau} & = f(x_{ik}, p), \quad k = 1, \cdots, K, \quad i = 1, \cdots, N - 1 \quad \rightarrow \text{Differential equation constarint} \\
    x_{i + 1, 0} & = \sum _{j = 0}^K l_j(1)x_{ij}, \quad i = 1, \cdots, N - 1 \quad \rightarrow \text{Continuity constraint}
\end{aligned}
\end{equation}

By enforcing the differential equation constraints at the collocation points, the continuous dynamic optimization problem is converted to a finite-dimensional algebraic optimization problem. This eliminates the need to explicitly integrate the ODEs during optimization, which is particularly advantageous for stiff systems where numerical integration is both computationally expensive and prone to instability. The objective function minimizes the error between the experimental and predicted values of the state at the measurement time points. Note that the measurement time points need not coincide with the finite element boundaries or the collocation points. The state at any measurement point can simply be evaluated using the polynomial approximation. A simple code example of estimating the parameters of Lotka-Volterra system is shown below

\begin{lstlisting}[style=pythonstyle]
    import numpy as np
    from scipy.integrate import odeint
    import pyomo.environ as pmo
    from pyomo.dae import ContinuousSet, DerivativeVar
    
    
    # Governing equations
    def LotkaVolterra(x, t):
        a, b, c, d = 2/3, 4/3, 1, 1
        return np.array([
            a * x[0] - b * x[0] * x[1],
            - c * x[1] + d * x[0] * x[1]
        ])

    # Generate data
    nx = 2 # dimensions of x
    t0, tf, dt = 1, 10, 0.1
    time_span = np.arange(t0, tf, dt)
    xinit = np.array([0.1, 0.2])
    solution = odeint(LotkaVolterra, xinit, time_span)
    
    # Initialize pyomo model
    model = pmo.ConcreteModel()
    model.t = ContinuousSet(initialize = time_span, bounds = (t0, tf))
    model.nx = pmo.RangeSet(0, nx - 1) # Dimensions of x
    
    model.a = pmo.Var(initialize = 0) # Parameter
    model.b = pmo.Var(initialize = 0) # Parameter
    model.c = pmo.Var(initialize = 0) # Parameter
    model.d = pmo.Var(initialize = 0) # Parameter
    
    # Define states and its derivatives
    model.x = pmo.Var(
        model.nx, model.t, 
        rule = lambda m, i, t : xinit[i]
    ) # states
    model.dxdt = DerivativeVar(model.x, wrt = model.t)
    
    # Fix the initial conditions
    for i in model.nx :
        model.x[i, model.t.first()].fix(xinit[i])
    
    # Governing differential equations as constraints
    @model.Constraint(model.nx, model.t)
    def _dxdt_rule(m, i, t):
        if t == m.t.first()  : return pmo.Constraint.Skip
        if i == 0 : 
            return (m.dxdt[i, t] == 
                    m.a * m.x[0, t] - m.b * m.x[0, t] * m.x[1, t])
        else : 
            return (m.dxdt[i, t] == 
                    - m.c * m.x[1, t] + m.d * m.x[0, t] * m.x[1, t])
    
    # Define objective function
    @model.Objective(sense = pmo.minimize)
    def objective_function(m):
        
        asum = 0
        points = 0
        for i in model.nx :
            for j, t in enumerate(time_span) : 
                asum += (m.x[i, t] - solution[j, i])**2
                points += 1
        
        return asum / points # MSE
    
    # Define collocation as discretization scheme
    discretizer = pmo.TransformationFactory('dae.collocation')
    discretizer.apply_to(
        model, wrt = model.t, nfe = 200, ncp = 4, 
        scheme = 'LAGRANGE-RADAU'
    )
    
    # Set solver options and solve the problem
    solver = pmo.SolverFactory('ipopt')
    solver.options['tol'] = 1e-8 # Tolerance
    solver.options['max_iter'] = 1000 # Max iterations
    solver.options['print_level'] = 5
    
    results = solver.solve(model, tee = True)
    p_opt = np.array([model.a.value, model.b.value, model.c.value, model.d.value])
\end{lstlisting}

\clearpage
\bibliographystyle{unsrtnat}
\bibliography{references}
\end{document}